\documentclass[prd,superscriptaddress,amsfonts,amssymb,amsmath,showpacs,onecolumn]{revtex4-2}
\usepackage{bm}
\usepackage{amsfonts}
\usepackage{amssymb}
\usepackage{accents}   
\usepackage{latexsym}
\usepackage{graphicx}
\usepackage{amsmath}
\usepackage{palatino}
\usepackage{mathpazo}
\usepackage{textcomp}
\usepackage{float}
\usepackage{booktabs}
\usepackage{dcolumn}
\usepackage{booktabs}
\usepackage{multirow}
\usepackage{hyperref}
\hypersetup{colorlinks,citecolor=blue}
\usepackage{amsmath}
\usepackage{xcolor}
\usepackage{orcidlink}
\usepackage{epsfig}
\usepackage{caption}
\usepackage{subcaption}
\usepackage{commath}
\usepackage{tensor}
\hypersetup{colorlinks,citecolor=blue}
\hypersetup{colorlinks=true,linkcolor=magenta,filecolor=magenta,    urlcolor=blue}

\def\be{\begin{equation}}
\def\ee{\end{equation}}
\def\bea{\begin{eqnarray}}
\def\eea{\end{eqnarray}}

\begin{document}

\title{Semi-Symmetric Metric Gravity: A Brief Overview}

\author{Himanshu Chaudhary}
\email{himanshu.chaudhary@stud.ubbcluj.ro} 
\affiliation{Department of Physics, Babeș-Bolyai University, Kogălniceanu Street, Cluj-Napoca, 400084, Romania,}
\author{Lehel Csillag}
\email{lehel.csillag@ubbcluj.ro} 
\affiliation{Department of Physics, Babeș-Bolyai University, Kogălniceanu Street, Cluj-Napoca, 400084, Romania,}
\affiliation{Department of Mathematics and Computer Science, Transilvania University,  Iuliu
 Maniu Street 50, Brașov 500091, Romania}
\author{Tiberiu Harko}
\email{tiberiu.harko@aira.astro.ro (Corresponding Author)}
\affiliation{Department of Physics, Babeș-Bolyai University, Kogălniceanu Street, Cluj-Napoca, 400084, Romania,}
\affiliation{Astronomical Observatory, 19 Cireșilor Street, Cluj-Napoca 400487, Romania,}

\begin{abstract}
We present a review of the Semi-Symmetric Metric Gravity (SSMG) theory, representing a geometric extension of standard general relativity, based on a connection introduced by Friedmann and Schouten in 1924. The semi-symmetric connection is a connection that generalizes the Levi-Civita one, by allowing for the presence of a simple form of the torsion, described in terms of a torsion vector. The Einstein field equations are postulated to have the same form as in standard general relativity, thus relating the Einstein tensor constructed with the help of the semi-symmetric connection, with the energy-momentum tensor. The inclusion of the torsion contributions in the field equations has intriguing cosmological implications, particularly during the late-time evolution of the Universe. Presumably, these effects also dominate under high-energy conditions, and thus SSMG could potentially address unresolved issues in General Relativity and cosmology, such as the initial singularity, inflation, or the $^7$Li problem of the Big-Bang Nucleosynthesis. The explicit presence of torsion in the field equations leads to the non-conservation of the energy-momentum tensor, which can be interpreted within the irreversible thermodynamics of open systems, as describing particle creation processes. We also review in detail the cosmological applications of the theory, and investigate the statistical tests for several models, by constraining the model parameters via comparison with several observational datasets.
\end{abstract}

\maketitle

\tableofcontents
\section{Introduction}
The advent of General Relativity \cite{Ein, Hilb}, a geometric theory of gravity, essentially based on Riemann geometry \cite{Riem}, had a profound effect not only on gravitational physics, but also on various branches of mathematics, and especially on differential geometry. The developments in mathematics also deeply influenced the understanding, and the approaches for the description of gravity, thus leading to a creative and dynamic interplay between mathematics and physics, the two fundamental sciences offering together the most fundamental tools for the investigation of the natural phenomena. The Riemann geometry, used by Einstein and Hilbert to build general relativity is metric, with the metric tensor $g_{\mu \nu}$ satisfying the condition $\overset{\circ}{\nabla} _\lambda g_{\mu \nu}=0$, where $\overset{\circ}{\nabla} _\lambda$ is the covariant derivative defined with the help of the Levi-Civita connection.

A few years after the field equations of general relativity were obtained, Weyl \cite{Weyl1, Weyl2,Weyl3} proposed an extension of Riemann geometry in which the metric condition is abandoned, and a new geometric concept, the nonmetricity $Q_{\lambda \mu \nu}$ is introduced. Weyl geometry is nonmetric, with the metric tensor satisfying the condition $\widetilde{\nabla}g_{\mu \nu}=Q_{\lambda \mu \nu}$, where $\widetilde{\nabla}$ is the covariant derivative defined with respect to the Weyl connection $\tensor{\widetilde{\Gamma}}{^\lambda _{\mu \nu}}$. 

Generalizations and applications of Weyl geometry, and of the related idea of conformal invariance were considered by Dirac \cite{Di1,Di2}, and by Penrose \cite{Pe1,Pe2}. An interesting approach to nonmetric geometries was developed by Schr\"{o}dinger \cite{Sch1,Sch2}, in which the length of vectors is conserved under autoparallel transport. For the physical applications of Schr\"{o}dinger geometry see \cite{Sch3,Sch4}. The role of Weyl geometry in physics, astrophysics and cosmology has been intensively investigated. In particular, it represents the mathematical foundation of the $f(Q)$ type generalized gravity theories, and of its extensions \cite{fQ1,fQ2,fQ3,fQ4, fQ5}. For a recent review of $f(Q)$ theories, see \cite{fQ6}.

Another remarkable development in mathematics, closely related to general relativity, was the introduction of the concept of torsion by Cartan \cite{Ca1,Ca2,Ca3,Ca4}. Cartan introduced the torsion as the antisymmetric part of an asymmetric affine connection, so that $\tensor{T}{^\lambda _{\mu \nu}}=\tensor{\Gamma}{ ^\lambda _{\nu \mu}}-\tensor{\Gamma}{ ^\lambda _{\mu \nu}}$. He also realized the tensor character of the torsion, and he developed the differential geometric formalism necessary to describe it. 

Cartan proposed a physical interpretation of the torsion of the space-time,  suggesting that it may be related to the intrinsic angular momentum (spin)
of matter, and that it vanishes in the matter-free regions of the spacetime. Cartan's ideas were forgotten later on, but they were reconsidered and extended in the 1960's, leading to the formulation of the so-called Einstein-Cartan theory  (see \cite{Hehl} for a detailed review of the early developments in this field), in which the spin of matter couples to a non-Riemannian geometric structure, namely, the torsion tensor. For discussions  on the cosmological applications and mathematical foundations of the Einstein-Cartan theory see \cite{EC1,EC2,Luz}.

From a cosmological point of view one of the attractive features of the Einstein-Cartan theory is the absence of the initial singularity, and the possibility of inflation in the early stages of the evolution of the Universe. 

A theory of gravity with a propagating massive vector field arising from a quadratic curvature invariant, in which the Einstein-Cartan formalism and a partial suppression of torsion lead to the absence of ghost and strong-coupling problems, was considered in \cite{Bark}.  By assuming a propagating torsion vector, one arrives at a purely gravitational origin of the Einstein-Proca models, and one could also constrain their parameter space. The reflection and transmission problem for a beam travelling across a spin-polarized target was considered in the non-relativistic limit in \cite{Bruno}. Deviations in the spin polarizations on the reflected and transmitted beams could distinguish Einstein-Cartan from general relativity, and, if detected, they would represent a strong evidence for a non-trivial spacetime torsion.

In the development of the physically related mathematical concepts an important role is played by the Weitzenb\"{o}ck spaces \cite{Weit}.
A Weitzenb\"{o}ck space is characterized by the geometric properties $\nabla _\lambda g_{\mu \nu}=0$, $\tensor{T}{^\lambda _{\mu \nu}}\neq 0$, and $\tensor{R}{^\lambda _{\sigma \mu \nu}}=0$, respectively, where $\tensor{R}{^\lambda _{\sigma \mu \nu}}$ is the Riemann curvature tensor.

Weitzenb\"{o}ck type geometries were used by Einstein \cite{EinW} for the unification of the electromagnetic and the gravitational fields in the framework of a teleparallel theory. As the basic geometric quantity in the teleparallel formulation of gravity the torsion tensor, generated by the tetrad fields, is adopted, with the curvature being replaced by the torsion. This description of the gravitational interaction is called the teleparallel equivalent of General Relativity (TEGR), and it was developed initially in \cite{TEGR1,TEGR2,TEGR3}. 

TEGR is also known as the $f(T )$ theory of gravity, where $T$ is the torsion scalar. The basic property of the $f(T )$ type theories of gravity is that torsion exactly cancels curvature, and therefore the curved space-time of general relativity reduces to a flat manifold. Moreover, instead of the metric the basic dynamical variables are the tetrad fields, and the gravitational Lagrangian density is constructed from the torsion scalar $T$ only. The corrections corresponding to higher energy scales can be represented  with the help of the higher order terms in torsion.

 The $f(T )$ theory has been extensively used in cosmology, for the explanation of inflation \cite{TEGR4}, of the late-time cosmic acceleration \cite{TEGR5}, and for obtaining  exact charged black hole solution, which contains, together with the monopole term, a quadrupole term, having its origin in the quadratic form $f(T )\propto  T^2$ \cite{TEGR6}. Matter and geometry couplings were considered in \cite{TEGR6a}. The $f(T )$ gravity theories have the major advantage that the gravitational field equations are of second order. For an in depth description of teleparallel theories and their applications see \cite{TEGR7,TEGR8, TEGR9}.

Gravitational theories in a Weyl-Cartan space-time, in which the Weitzenb\"{o}ck condition of the vanishing of the sum of the curvature and torsion scalar is also imposed, where considered in \cite{WC1,WC2}, with a kinetic term for the torsion also included in the gravitational action. The field equations of the theory, obtained by varying the action with respect to the metric tensor, lead to a complete description of the gravitational field in terms of two fields, the Weyl vector and the torsion, respectively. Within this theory a large variety of dynamic evolutions can be obtained, ranging from inflationary/accelerated expansions to non-inflationary behaviors.

In 1924, at around the same time when Cartan was developing the concept of torsion, by also considering its applications to general relativity, Friedmann and Schouten \cite{FS} have introduced a new geometrical concept, the semi-symmetric transport, in which they assumed that $S^{..\nu}_{\lambda \mu}=(1/2)\left(\Gamma ^\nu_{\lambda \mu}-\Gamma ^\nu_{\mu \lambda}\right)=S_{[\lambda}A_{\mu]}^\nu$, where $A_\lambda ^\nu=1, \lambda =\nu$, and $A_\lambda ^\nu=0, \lambda \neq \nu$. Friedmann and Schouten did not mention in their work the previous investigations by Cartan, or the concept of torsion. In their work, they also mentioned that a connection can be completely characterized by the so-called covariant differential quotient (nowadays known as non-metricity) and the covariant antisymmetric parameters (nowadays known as the components of torsion). This characterization in modern terminology is known as the post-Riemannian decomposition. Although mostly of mathematical interest, they also obtained autoparallel equations of a semi-symmetric connection. As for physical applications, they considered a special affine connection with semi-symmetric type of torsion, and claimed it could perhaps incorporate electromagnetism into the geometric framework, although this was not explicitly shown in the article. 

The connection introduced by Friedmann and Schouten, one hundred years ago, achieved prominent mathematical applications. The work of Hayden \cite{Hayden1932}, in which he studied submanifolds with torsion, raised interest in finding submanifolds with semi-symmetric metric connections. Such submanifolds have been found and presented for the first time in \cite{Amur1978}. The foundational work of  Kentaro Yano \cite{Yano1970}, in which he computed for the first time the curvature tensors of the semi-symmetric metric connection in a coordinate-free manner, raised the interest of mathematicians for this connection. Since then, many studies have been conducted, both assuming metric compatibility \cite{EstonianAcademy2008,Balgeshir2021,Chaturvedi2014, Guvenc2024} , and metric incompatibility \cite{De2025} on statistical manifolds \cite{Yildirim2022}. Non-metric semi-symmetric connections have also been lifted to the tangent bundle \cite{De2023}, giving preliminary steps to the application of semi-symmetric connections in Finsler geometry, whose central object is the slit tangent bundle $TM \setminus \{0\}$.  It is still an open problem to find a complex structure on a Kähler manifold with respect to a newly constructed connection from a semi-symmetric metric connection \cite{Mihai2017}.

General relativity is an extremely successful physical theory,  which gives an almost perfect description of the gravitational physics at the level of the Solar system. However, presently the theory of general relativity faces a number of important challenges. First of all, from a fundamental theoretical point of view, it cannot explain the quantum properties of the gravitational interaction, or provide a consistent view on the gravitational effects at a quantum scale. Secondly, gravitational collapse can result in the formation of singularities, due to the geodesic incompleteness, and under some specific assumptions on the matter energy-momentum tensor. An important consequence of these results is the appearance of cosmological singularities during the Big Bang, and the existence of the black holes.

An important challenge for general relativity did appear after the discovery of the late-time cosmic accelerated expansion of the Universe. Important evidence provided by the observations of type Ia supernovae (SN Ia) \cite{acc1,acc2,acc3}, large-scale structure observations, as well as the determinations of the Cosmic Microwave
Background (CMB) anisotropies by the Wilkinson Microwave Anisotropy Probe (WMAP) \cite{WMAP}, and by the Planck satellite \cite{Planck} strongly pointed out towards the limitation of general relativity's theoretical potential to describe and understand the evolution of the Universe at cosmological scales during its recent phases. To explain the accelerating expansion of the Universe, inferred from the observations of the luminosity distance of the SN Ia, one must extend general relativity by adding new terms in the Einstein field equations, representing either a cosmological constant, or a dark energy. The confrontation of the Planck satellite observations with the SN Ia data has led to different predictions on the present day numerical values of the basic cosmological parameters.  

The
failure of general relativity as confronted with the latest observations led to the necessity of exploring alternative theories of gravity,  and of introducing new
terms or physical components in the gravitational field equations, such as dark energy, and dark matter, or the addition of a cosmological constant $\Lambda$ in the Hilbert-Einstein action.  To explain the observations within a theoretical framework,  the $\Lambda$CDM ($\Lambda$ Cold Dark Matter) model was developed, which is
based on the reconsideration in the gravitational field equations of the cosmological term $\Lambda$, proposed by Einstein in 1917 \cite{EinL} to obtain a static cosmological model of the Universe. After the discovery of the expansion of the Universe, Einstein rejected the possibility of the existence of $\Lambda$, and suggested its removal from the gravitational field equations.

The $\Lambda$CDM model provides a very good description of the observational data at low redshifts, and thus it has been adopted as the standard cosmological paradigm of the present times. Nevertheless, the $\Lambda$CDM model is confronted with an important problem associated to its basis. No satisfactory explanation of the physical or geometrical nature  of $\Lambda$ is presently known, and thus the theoretical foundations of the $\Lambda$CDM model are rather questionable.

The $\Lambda$CDM model must also face with several other challenges, which follow from the increase in the accuracy of the cosmological measurements and observations. An important problem present day cosmology faces is the significant difference between the expansion rate of the Universe as determined from the Planck satellite observations, and the numerical values obtained from the local  determinations \cite{HT1}. This discrepancy is usually called as the Hubble tension \cite{HT2,HT3}. As measured by the Planck satellite, the Hubble constant $H_0$ has the numerical value of $66.93 \pm  0.62$ km/ s/ Mpc \cite{HT4}. From the  SH0ES collaboration the value of
$73.24 \pm  1.74$ km/ s/ Mpc \cite{HT4} is obtained for $H_0$. There is a difference of more than 3$\sigma$ \cite{HT4} between these values. If indeed the Hubble tension does exist,
it strongly points out towards the demand of investigating alternative gravitational theories, and generalizing, or even completely
superseding, the $\Lambda$CDM model. 

The $\Lambda$CDM paradigm naturally  incorporates the Big Bang concept, which is based on three fundamental observational facts, the Hubble expansion of the Universe, the Cosmic Microwave Background Radiation (CMBR), and the Big Bang Nucleosynthesis (BBN), respectively. Because of the precise determinations of the baryon-to-photon ratio obtained
from the studies of the anisotropies of CMBR, the Standard BBN is a parameter-free theory. The theoretically computed abundances of light elements formed during primordial nucleosynthesis and those determined from observations are in good agreement throughout a range of nine orders of magnitude.

However,  there is still an important disparity between theory and observation in the 
of $^7$Li abundance, which is overestimated by a factor of $\sim 2.5$ when calculated theoretically. This problem is called the cosmological lithium problem \cite{Li1}. In order to solve it, going beyond the standard model of cosmology may be necessary \cite{Li2}.

The late-time cosmic acceleration, as well as other observational evidence including the problem of the structure formation, has led to the question wether general relativity is the correct relativistic theory of gravitation. Presently, general relativity is facing many challenges, like, for example, the difficulty of explaining a large number of observations, the incompatibility with quantum mechanics, 
 and the lack of uniqueness, respectively.
  
 These observational/theoretical facts naturally point towards the necessity for new gravitational physics, and for the formulation of a new fundamental theory of gravity.
 
 An ideal testing ground for general relativity is represented by cosmology, through the investigation of the cause, and characteristics of the recent cosmic acceleration. 
A promising approach to solve this problem is to consider that at large, astrophysical or cosmological, scales, general relativity breaks down, and a more general theory describes the gravitational interaction. 

The physical or mathematical motivations for the extensions of general relativity also include the possibility of a realistic description of the gravitational physics near curvature singularities, as well as obtaining some simple first order effective approximations for the elusive quantum theory of the gravitational fields. Hence, by taking into account the present day situation in gravitational physics, one needs to consider, or reconsider, alternative approaches to gravity, which could at least give a better description of the observational data.

It is the goal of the present work to consider in detail the general physical, astrophysical and cosmological implications of the Friedmann-Schouten geometry \cite{FS}, which was considered for a first time in a gravitational context in \cite{CSM}. After a brief review of the mathematical formalism, the gravitational field equations are written down. They generalize the standard Einstein equations through the inclusion of the contributions of the torsion vector. 

The Newtonian limit of the theory leads to the generalized Poisson equation, which explicitly includes the effects of the torsion, and allow us to obtain the corrections to the Newtonian potential, and to the Newtonian force. The applications of these results to the solution of the dark matter problem are also briefly discussed. 

The cosmological implications of the theory are investigated in detail, for a flat and homogeneous geometry. After obtaining the Friedmann equations in a general form, we consider the cosmological tests of the theory by analyzing two specific models, obtained by imposing two particular equations of state on the effective dark energy and pressure, which appear as an extra contribution in the Friedmann equations. The theoretical predictions are then compared to a set of Cosmic Chronometer, supernovae, and Baryon Acoustic Oscillations (BAO) measurements. We also discuss in detail the behavior of the cosmographic parameters, and a full comparison with the predictions of the standard $\Lambda$CDM model is also provided. 

Our results suggest that the Semi-Symmetric Metric Gravity cosmology has the potential of explaining cosmological dynamics without the need of introducing a cosmological constant in the theory, by means of a geometrically generated dark energy term. 

The present paper is organized as follows.  The field equations of the theory, the conservation equation, the Newtonian limit, and the thermodynamic interpretation of the matter non-conservation are discussed in Section~\ref{sect1}. Two particular cosmological models, as well as their cosmological tests, are presented in Section~\ref{sect2}. Finally, we discuss and conclude our results in Section~\ref{sect3}.     

\section{Semi-Symmetric Metric Gravity}\label{sect1}

In this section, we review the recently developed gravitational theory with arguably the simplest form of torsion, completely determined by a vectorial degree of freedom. Instead of focusing on the mathematical aspects and the coordinate-free approach developed in \cite{CSM}, we derive everything in coordinates.
\subsection{Semi-Symmetric Metric geometry}
The Semi-Symmetric metric connection belongs to the class of metric-compatible connections with torsion. This connection was first introduced by Friedmann and Schouten in 1924 \cite{FS}, and its connection coefficient functions are given by
\begin{equation}\label{recoverthis}
     \tensor{{\Gamma}}{^\mu _\nu _\rho}= \tensor{\gamma}{^\mu _\nu _\rho} - \pi^{\mu} g_{\rho \nu} + \pi_{\nu} \delta^{\mu}_{\rho},
\end{equation}
where $\pi_\nu$ is a four-vector and $\tensor{\gamma}{^\mu _\nu _\rho}$ are the Christoffel symbols defined as
\begin{equation}
     \tensor{\gamma}{^\mu _\nu _\rho}=\frac{1}{2} g^{\mu \lambda} \left( \partial_{\nu} g_{\rho \lambda}+ \partial_{\rho} g_{\lambda \nu} - \partial_{\lambda} g_{\nu \rho} \right).
\end{equation}
Hence, one can observe that the Semi-Symmetric metric connection differs by the Levi-Civita connection by a vectorial degree of freedom, $\pi$. This specific type of torsion has also been considered in a recent extension of the singularity theorems to geometries with torsion \cite{vanDeVenn2024}. Its simplicity allows to draw new conclusions, which are not obvious at all when the torsion tensor has non-diagonal components as well. Moreover, such a type of torsion can be naturally obtained as a solution to the connection field equations in interacting hyperhydrodynamical models \cite{Iosifidis2024}. Some bounds on the torsion vector $\pi$ have been obtained in \cite{Kranas2019}, by considering a so-called steady-state approximation (for more details consult \cite{Kranas2019}). The theory proposed has also been investigated from a dynamical systems perspective \cite{Barrow2019}.

Recall that in non-metric torsion-free geometries, there is also a special type of non-metricity, which is fully determined by a vectorial degree of freedom. This is the Weyl geometry, given by
\begin{equation}
    \nabla_{\mu} g_{\nu \rho}=w_{\mu} g_{\nu \rho}.
\end{equation}
In metric-affine $f(R)$ theories, there is a duality between the Semi-Symmetric type of torsion and Weyl-type non-metricity \cite{Iosifidis2019}. Note that the connection coefficient functions of the Semi-Symmetric metric connection can be obtained as a special case from the general post-Riemannian expansion of a metric-compatible torsionful connection, which reads
\begin{equation}\label{connection}
    \tensor{{\Gamma}}{^\mu _\nu _\rho}=\tensor{\gamma}{^\mu _\nu _\rho}
    - \frac{1}{2}g^{\lambda \mu}(T_{\rho \nu \lambda}+T_{\nu \rho \lambda}- T_{\lambda \rho \nu}),
\end{equation}
where the torsion tensor is twice the antisymmetric part of the connection coefficient functions
\begin{equation}
    \tensor{T}{^\mu_{\nu \rho}}=2 \tensor{\Gamma}{^\mu _{[\rho \nu]}}.
\end{equation}

To recover \eqref{recoverthis}, the Semi-Symmetric type of torsion
\begin{equation}\label{semisymmetric}
    \tensor{T}{^\mu_\nu _\rho}=\pi_{\rho} \delta^{\mu}_{\nu} - \pi_{\nu} \delta^{\mu}_{\rho}
\end{equation}
has to be substituted in \eqref{connection}. It is easily seen that in this special case, the torsion is completely determined by a four-vector as well. Hence, in the literature, this type of torsion is sometimes called \textit{vectorial torsion} \cite{agricola2004,agricola2016}. Arguably, this is the simplest form of torsion one could consider, as it has only  diagonal components. The Riemann tensor of a general affine connection is given by
\begin{equation}
\begin{aligned}
    R\tensor{}{^\mu _\nu _\rho _\sigma}=\tensor{\Gamma}{^\lambda _\nu _\sigma} \tensor{\Gamma}{^\mu _\lambda _\rho} - \tensor{\Gamma}{^\lambda _\nu _\rho} \tensor{\Gamma}{^\mu_\lambda _\sigma}
   + \partial_{\rho} \tensor{\Gamma}{^\mu _\nu _\sigma} - \partial_{\sigma} \tensor{\Gamma}{^\mu _\nu _\rho},
   \end{aligned}
\end{equation}
while the Ricci tensor and scalar are its contractions
\begin{equation}
    R_{\nu \sigma}=R\tensor{}{^\mu _\nu _\mu _\sigma}, \; \; R=g^{\nu \sigma} R_{\nu \sigma}.
\end{equation}
To obtain a compact formula using the post-Riemannian expansion, let us recall the distortion tensor \cite{CSM}
\begin{equation}
    \tensor{N}{^\mu _\nu _\rho}= \tensor{\Gamma}{^\mu _\nu _\rho} - \tensor{\gamma}{^\mu _\nu _\rho}.
\end{equation}
The Ricci tensor can then be expressed as
\begin{equation}
\begin{aligned}
    R_{\nu \sigma}= \overset{\circ}{R}_{\nu \sigma}+ \accentset{\circ}{\nabla}_{\alpha} \tensor{N}{^\alpha _\nu _\sigma} - \accentset{\circ}{\nabla}_{\sigma} \tensor{N}{^\alpha  _\nu  _\alpha} 
    + \tensor{N}{^\alpha _\rho _\alpha} \tensor{N}{^\rho _\nu _\sigma} - \tensor{N}{^\alpha  _\rho _\sigma} \tensor{N}{^\rho _\nu _\alpha},
\end{aligned}
\end{equation}
where $\overset{\circ}{R}_{\nu \sigma}$ denotes the Ricci tensor of the Levi-Civita connection, and $\overset{\circ}{\nabla}$ denotes the Levi-Civita covariant derivative.
In the case of the Semi-Symmetric metric connection, for the distortion tensor we have
\begin{equation}
    \tensor{N}{^\mu _\nu _\rho}=-\pi^\mu g_{\nu \rho} + \pi_{\nu} \delta^{\mu}_{\rho}.
\end{equation}
Consequently, we get
\begin{equation}
\begin{aligned}
    R_{\nu \sigma}= \accentset{\circ}{R}_{\nu \sigma}-  g_{\nu \sigma} \accentset{\circ}{\nabla}_{\alpha} \pi^{\alpha}+ \accentset{\circ}{\nabla}_{\sigma} \pi_\nu -3 \accentset{\circ}{\nabla}_{\sigma} \pi_{\nu}
    -3 \pi_{\rho} \pi^{\rho} g_{\nu \sigma} +3 \pi_{\sigma }\pi_{\nu}- \pi_\nu \pi_\sigma  + g_{\nu \sigma} \pi^\alpha \pi_\alpha,
\end{aligned}
\end{equation}
or equivalently
\begin{equation}\label{Riccitensor}
    R_{\nu \sigma}=\accentset{\circ}{R}_{\nu \sigma}- g_{\nu \sigma} \accentset{\circ}{\nabla}_{\alpha} \pi^{\alpha} -2 \accentset{\circ}{\nabla}_{\sigma }\pi_{\nu} -2\pi_\rho \pi^\rho g_{\nu \sigma} + 2\pi_\nu \pi_\sigma.
\end{equation}
The Ricci scalar is readily obtained 
\begin{equation}\label{Ricciscalar}
    R=\accentset{\circ}{R}-6 \accentset{\circ}{\nabla}_{\alpha} \pi^{\alpha} -6 \pi_\alpha \pi^\alpha.
\end{equation}
\subsection{The gravitational field equations}
We postulate that the field equations take the form \cite{CSM}
\begin{equation}
    R_{(\nu \sigma)}- \frac{1}{2} g_{\nu \sigma}R= 8 \pi T_{\nu \sigma}.
\end{equation}
Substituting the post-Riemannian expansions \eqref{Riccitensor}, \eqref{Ricciscalar} yields
\begin{eqnarray}
    \accentset{\circ}{R}_{\nu \sigma} - \frac{1}{2} g_{\nu \sigma} \accentset{\circ}{R}- g_{\nu \sigma} \accentset{\circ}{\nabla}_{\alpha} \pi^{\alpha}
    - \accentset{\circ}{\nabla}_{\sigma} \pi_{\nu} &-& \accentset{\circ}{\nabla}_{\nu} \pi_{\sigma} 
    - 2 \pi_\rho \pi^\rho g_{\nu \sigma} + 2 \pi_\nu \pi_\sigma  \nonumber  + 3 g_{\nu \sigma} \accentset{\circ}{\nabla}_{\alpha} \pi^\alpha
    + 3 g_{\nu \sigma} \pi_\alpha \pi^\alpha  = 8 \pi T_{\nu \sigma}.
\end{eqnarray}
The above equation can be separated to directly see the contributions of the torsion vector $\pi_\rho$ as
\begin{equation}\label{feq}
\begin{aligned}
   \accentset{\circ}{R}_{\nu \sigma}-\frac{1}{2} g_{\nu \sigma} \accentset{\circ}{R} +2 g_{\nu \sigma} \accentset{\circ}{\nabla}_{\alpha} \pi^{\alpha} - \accentset{\circ}{\nabla}_{\sigma} \pi_{\nu} - \accentset{\circ}{\nabla}_{\nu} \pi_{\sigma}
    + \pi_{\rho} \pi^{\rho} g_{\nu \sigma} +2 \pi_{\nu} \pi_{\sigma}= 8 \pi T_{\nu \sigma}.
\end{aligned}
\end{equation}
By taking the trace of Eq.~(\ref{feq}) we obtain
\begin{equation}
-\accentset{\circ}{R}+6\accentset{\circ}{\nabla}_{\alpha }\pi ^{\alpha }+6\pi _{\rho }\pi
^{\rho }=8\pi T,
\end{equation}
and thus we obtain for the field equations the alternative form
\begin{equation}\label{Rnusi}
\accentset{\circ}{R}_{\nu \sigma }=8\pi \left( T_{\nu \sigma }-\frac{1}{2}Tg_{\nu
\sigma }\right) +\accentset{\circ}{\nabla}_{\alpha }\pi ^{\alpha }g_{\nu \sigma
}+2\pi _{\rho }\pi ^{\rho }g_{\nu \sigma }+\accentset{\circ}{\nabla}_{\sigma }\pi
_{\nu }+\accentset{\circ}{\nabla}_{\nu }\pi _{\sigma }-2\pi _{\nu }\pi _{\sigma }.
\end{equation}
\subsubsection{The divergence of the matter energy-momentum tensor}
In the mixed representation the gravitational field equations take the form
\begin{equation}
\accentset{\circ}{R}_{\nu }^{\sigma }-\frac{1}{2}\delta _{\nu }^{\sigma }\accentset{\circ}{R}%
+2\delta _{\nu }^{\sigma }\accentset{\circ}{\nabla}_{\alpha }\pi ^{\alpha }-%
\accentset{\circ}{\nabla}^{\sigma }\pi _{\nu }-\accentset{\circ}{\nabla}_{\nu }\pi ^{\sigma
}+\pi _{\rho }\pi ^{\rho }\delta _{\nu }^{\sigma }+2\pi _{\nu }\pi ^{\sigma
}=8\pi T_{\nu }^{\sigma }  \label{feqm}
\end{equation}
By taking the divergence of Eq. (\ref{feqm}), and since $\accentset{\circ}{\nabla}%
_{\sigma }\left( \accentset{\circ}{R}_{\nu }^{\sigma }-\frac{1}{2}\delta _{\nu
}^{\sigma }\accentset{\circ}{R}\right) \equiv 0$, we obtain the relation
\begin{equation}\label{16}
    - \accentset{\circ}{\nabla}_{\sigma } \accentset{\circ}{\nabla}^{\sigma } \pi _{\nu }
    + \accentset{\circ}{\nabla}_{\nu } \accentset{\circ}{\nabla}_{\sigma } \pi^{\sigma }
    + \left( \accentset{\circ}{\nabla}_{\nu } \accentset{\circ}{\nabla}_{\sigma }
    - \accentset{\circ}{\nabla}_{\sigma } \accentset{\circ}{\nabla}_{\nu } \right) \pi^{\sigma } + \pi^{\sigma } \accentset{\circ}{\nabla}_{\nu } \pi_{\sigma } + \pi_{\sigma } \accentset{\circ}{\nabla}_{\nu } \pi^{\sigma } + 2 \pi^{\sigma } \accentset{\circ}{\nabla}_{\sigma } \pi_{\nu }
    + 2 \pi_{\nu } \accentset{\circ}{\nabla}_{\sigma } \pi^{\sigma }
    = 8 \pi \accentset{\circ}{\nabla}_{\sigma } T_{\nu }^{\sigma }.
\end{equation}
The Riemann curvature tensor is defined according to
\begin{equation}
\left( \accentset{\circ}{\nabla}_{\nu }\accentset{\circ}{\nabla}_{\sigma }-\accentset{\circ}{\nabla}%
_{\sigma }\accentset{\circ}{\nabla}_{\nu }\right) \pi ^{\lambda }=-\pi ^{\alpha }%
\accentset{\circ}{R}_{\alpha \nu \sigma }^{\lambda }.
\end{equation}%
Contracting with $\lambda =\sigma $ we obtain
\begin{equation}\label{18}
\left( \accentset{\circ}{\nabla}_{\nu }\accentset{\circ}{\nabla}_{\sigma }-\accentset{\circ}{\nabla}%
_{\sigma }\accentset{\circ}{\nabla}_{\nu }\right) \pi ^{\sigma }=-\pi ^{\alpha }%
\accentset{\circ}{R}_{\alpha \nu \sigma }^{\sigma }=\pi ^{\alpha }\accentset{\circ}{R}%
_{\alpha \sigma \nu }^{\sigma }=\pi ^{\alpha }\accentset{\circ}{R}_{\alpha \nu }=\pi
^{\sigma }\accentset{\circ}{R}_{\nu \sigma }.
\end{equation}
From Eq. (\ref{Rnusi}) we find
\begin{equation}\label{19}
\pi ^{\sigma }\accentset{\circ}{R}_{\nu \sigma }=8\pi \left( \pi ^{\sigma }T_{\nu
\sigma }-\frac{1}{2}T\pi _{\nu }\right) +\pi _{\nu }\accentset{\circ}{\nabla}%
_{\sigma }\pi ^{\sigma }+\pi ^{\sigma }\accentset{\circ}{\nabla}_{\sigma }\pi _{\nu
}+\pi ^{\sigma }\accentset{\circ}{\nabla}_{\nu }\pi _{\sigma }.
\end{equation}
Hence, by substituting Eqs.~(\ref{19}) and (\ref{18}) into Eq.~(\ref{16}), we obtain the expression of the divergence of the matter energy-momentum tensor as
\begin{eqnarray}\label{23}
    8 \pi \accentset{\circ}{\nabla}_{\sigma } T_{\nu }^{\sigma } 
    &=&  - \accentset{\circ}{\nabla}_{\sigma } \accentset{\circ}{\nabla}^{\sigma } \pi_{\nu } 
    + \accentset{\circ}{\nabla}_{\nu } \accentset{\circ}{\nabla}_{\sigma } \pi^{\sigma } + 8 \pi \left( \pi^{\sigma } T_{\nu \sigma } - \frac{1}{2} T \pi_{\nu } \right) \nonumber + 3 \pi_{\nu } \accentset{\circ}{\nabla}_{\sigma } \pi^{\sigma } 
    + 3 \pi^{\sigma } \accentset{\circ}{\nabla}_{\sigma } \pi_{\nu } 
    + 3 \pi^{\sigma } \accentset{\circ}{\nabla}_{\nu } \pi_{\sigma } 
    \equiv  8 \pi f_{\nu }.
\end{eqnarray}
By imposing the condition $8\pi \accentset{\circ}{\nabla}_{\sigma }T_{\nu }^{\sigma }\equiv 0$, $f_\nu=0$, Eq.~(\ref{23}) provides an evolution equation for the torsion vector $\pi_\mu$. However, nonconservative models with $8\pi \accentset{\circ}{\nabla}_{\sigma }T_{\nu }^{\sigma }\neq 0$ can also be constructed in the framework of the present theory.
\subsection{Semi-Symmetric Einstein manifolds}
Before moving to the cosmological applications of  Semi-Symmetric Metric Gravity theory, we present a recent result on the existence of generalized Einstein manifolds with torsion. As it is well known, in the Riemannian setting, Einstein manifolds provide solutions to the vacuum field equations with a cosmological constant $\lambda$. They are characterized by the property
\begin{equation}\label{einsteinmanifold}
    \overset{\circ}{R}_{\mu \nu}=\lambda g_{\mu \nu},
\end{equation}
where $\lambda$ is a constant and $\overset{\circ}{R}_{\mu \nu}$ is the Ricci tensor of the Levi-Civita connection. Hence, in the Riemannian setting, condition \eqref{einsteinmanifold} is a system of partial differential equations for the metric $g_{\mu \nu}$. This system of partial differential equations has been thoroughly studied, and exact solutions were found \cite{Besse1987}. However, most of the solutions are static, meaning that the metric does not depend on time. An interesting generalization of Einstein manifolds to non-Riemannian geometries, in which torsion and non-metricity are present was proposed by Klemm and Ravera \cite{raveratorsionnonmetricity}. In this work, a generalized Einstein manifold is defined by
\begin{equation}\label{Einsteingeneralized}
    R_{(\mu \nu)}=\lambda g_{\mu \nu},
\end{equation}
where $R_{\mu \nu}$ is the Ricci tensor of the full connection and $\lambda$ is a function, not necessarily a constant. Note that in general this is not symmetric, and we do not put any constraint on the antisymmetric part of it, for the sake of this definition. As noted in \cite{csillag2024schrodingerconnectionsmathematicalfoundations}, this equation basically gives a constraint equation for the connection, since the Ricci tensor is not necessarily described solely by the metric in this generalized setting. Consequently, given a fixed metric $g_{\mu \nu}$, Eq. (\ref{Einsteingeneralized}) is a system of partial differential equations for the components of the connection. We present an example of a generalized Einstein manifold with torsion, but first, let us mention two possibilities of solving the constraint equation \eqref{Einsteingeneralized}
\begin{enumerate}
    \item Considering a metric, which is Einstein, in the sense that 
    \begin{equation}
        \overset{\circ}{R}_{\mu \nu}=\lambda g_{\mu \nu}
    \end{equation}
    is satisfied.
    \item Considering a metric, which is not Einstein, i.e.
    \begin{equation}
        \overset{\circ}{R}_{\mu \nu}=\lambda g_{\mu \nu}
    \end{equation}
    is not satisfied.
\end{enumerate}
Hence, this generalization, could allow for a non-static metric, if there exists a connection, which solves Eq. \eqref{Einsteingeneralized}.
 In the following, we will present the recently obtained results in this direction for the Semi-Symmetric metric connection. In \cite{csillag2024schrodingerconnectionsmathematicalfoundations} it is shown that condition \eqref{Einsteingeneralized} for a Semi-Symmetric connection is equivalent to
\begin{equation}\label{einsteintorsion}
    \overset{\circ}{R}_{\mu \nu} - \overset{\circ}{\nabla}_{\mu} \pi_{\nu} - \overset{\circ}{\nabla}_{\nu} \pi_{\mu}+ 2\pi_\nu \pi_\mu + \frac{1}{2} g_{\mu \nu} \overset{\circ}{\nabla}_{\lambda} \pi^{\lambda} - \frac{1}{2} g_{\mu \nu} \pi_\lambda \pi^\lambda=\frac{1}{4} g_{\mu \nu} \overset{\circ}{R}.
\end{equation}
In addition, if the metric $g_{\mu \nu}$ is an Einstein metric, the equation can be simplified to
\begin{equation}
    - \overset{\circ}{\nabla}_{\mu} \pi_{\nu} - \overset{\circ}{\nabla}_{\nu} \pi_{\mu}+ 2\pi_\nu \pi_\mu + \frac{1}{2} g_{\mu \nu} \overset{\circ}{\nabla}_{\lambda} \pi^{\lambda} - \frac{1}{2} g_{\mu \nu} \pi_\lambda \pi^\lambda=0,
\end{equation}
which gives an alternative evolution equation for the torsion vector $\pi$, in a given background Einstein metric $g_{\mu \nu}$.
\paragraph{Cosmology of the Semi-Symmetric Einstein manifolds.} For a Semi-Symmetric metric connection, considering a flat Friedmann-Lemaître-Robertson-Walker (FLRW) type Universe, described by the metric
\begin{equation}
    ds^2=dt^2-a(t)^2 \delta_{ij} dx^i dx^j,
\end{equation}
assuming that $\frac{\dot a(t)}{a(t)}=\text{const}$, in accordance with the cosmological principle \cite{TSAMPARLIS197927}, the torsion vector has only one non-vanishing component
\begin{equation}
    \pi_\mu=(\psi(t),0,0,0).
\end{equation}
Consequently, in this non-static background metric, Eq.~\eqref{einsteintorsion} becomes a differential equation for $\psi(t)$. In \cite{csillag2024schrodingerconnectionsmathematicalfoundations} it is shown that it admits an analytical solution, given by
\begin{equation}
 \psi(t)=-\frac{H_0 e^{H_0(t_0+t)}}{e^{H_0(t_0+t)}-1},
\end{equation}
where $t_0$ is an arbitrary integration constant fixed by the initial condition $\psi(0)$. Hence, Semi-Symmetric connections give a first explicit example of a generalized non-static Einstein manifold with torsion.
\subsubsection{The Newtonian limit}
In order to obtain the Newtonian limit of the Semi-Symmetric Metric Gravity theory we assume, similarly to the standard general relativistic case, that in the limit of small velocities and weak gravitational fields, the expression of the $g_{00}$ metric tensor component can be approximated as
\begin{equation}
g_{00}(r)=1+2\phi (r),
\end{equation}
and $g^{00}\approx 1$, where $\phi$ is the Newtonian potential. Moreover, $\sqrt{-g}\approx 1$. In the following we will neglect the time dependence of all physical and geometrical quantities. Furthermore, for the components of the energy-momentum tensor we adopt the expressions $T_\nu^\sigma =\rho u_\nu u^\sigma$, where$\rho$ is the matter energy density. For the case of the slow motion, we can neglect in the expression of the four-velocity all space components, and keep only the time component. Hence, the components of the four velocity are $u^\mu=(1,0,0,0)$, giving for the only non-zero component of the matter energy-momentum tensor the expression $T_0^0=\rho(r)$ and $T=\rho (r)$, respectively. We also represent the torsion vector as $\pi ^\rho\approx \pi _\rho =\left(\pi _0(r), \vec{\Pi}(r)\right)$, where $\vec{\Pi}(r)=\left(\pi_1(r),\pi_2(r),\pi_3(r)\right)$.
The only field equation of interest is thus Eq.~(\ref{Rnusi}) for $\nu =\sigma =0$, which leads to
\begin{equation}
\accentset{\circ}{R}_{0}^{0}=4\pi \rho +\accentset{\circ}{\nabla}_{\alpha }\pi ^{\alpha
}\delta _{0}^{0}+2\pi _{\rho }\pi ^{\rho }\delta _{0}^{0}+2\accentset{\circ}{\nabla}%
^{0}\pi _{0}-2\pi ^{0}\pi _{0}.
\end{equation}
To obtain the expression of $\accentset{\circ}{R}_{0}^{0}$, we neglect the products of the terms containing the products \ of the Christoffel symbols. Moreover, the terms containing the time derivative can also be neglected. Hence, the only non-zero terms in the expression of $\accentset{\circ}{R}_{0}^{0}$ are $%
\accentset{\circ}{R}_{00}\approx \accentset{\circ}{R}_{0}^{0}=\partial \accentset{\circ}{\Gamma}%
_{00}^{\mu }/\partial x^{\mu }$. As for the expressions of the Christoffel
symbols we obtain $\accentset{\circ}{\Gamma}_{00}^{\mu }\approx -(1/2)g^{\mu \nu
}\partial g_{00}/\partial x^{\nu }=\partial \phi /\partial x^{\mu }$. Hence, we obtain $\accentset{\circ}{R}_{0}^{0}\approx \Delta \phi $, where $\Delta $ is the
three-dimensional Laplace operator defined in the Newtonian-Galilean geometry. For the term $\accentset{\circ}{\nabla}_{\alpha }\pi ^{\alpha }$ we find
\begin{equation}
\accentset{\circ}{\nabla}_{\alpha }\pi ^{\alpha }=\frac{1}{\sqrt{-g}}\frac{\partial
}{\partial x^{\alpha }}\left( \sqrt{-g}\pi ^{\alpha }\right) \approx \frac{%
\partial \pi ^{\alpha }}{\partial x^{\alpha }}=\vec{\nabla}\cdot \vec{\Pi},
\end{equation}%
where by $\vec{\nabla}$ we have denoted the three-dimensional divergence operator. For the term $\accentset{\circ}{\nabla}^{0}\pi _{0}$ we obtain
\begin{equation}
\accentset{\circ}{\nabla}^{0}\pi _{0}\approx \accentset{\circ}{\nabla}_{0}\pi _{0}=\frac{%
\partial \pi _{0}}{\partial x^{0}}-\accentset{\circ}{\Gamma}_{00}^{\mu }\pi _{\mu
}\approx \frac{1}{2}g^{\mu \nu }\frac{\partial g_{00}}{\partial x^{\nu }}\pi
_{\mu }=\vec{\nabla}\phi \cdot \vec{\Pi}.
\end{equation}

Hence, we obtain the generalized Poisson equation in the Semi-Symmetric
Metric Gravity in the form
\begin{equation}\label{GP}
\Delta \phi =4\pi \rho +\vec{\nabla}\phi \cdot \vec{\Pi}+\vec{\nabla}\cdot
\vec{\Pi}+2\vec{\Pi}^{2}.
\end{equation}

As a simple illustrative example of the effect of the torsion on the
gravitational field we consider the case in which the torsion vector has
only one nonzero component, which, for simplicity, we assume to be constant.
Hence  $\vec{\Pi}=\left( \Pi _{r}(r),0,0\right) $, with $\Pi _{r}=$
constant. In vacuum, with $\rho =0$, in spherical symmetry, the generalized
Poisson equation (\ref{GP}) takes then the form
\begin{equation}
\frac{1}{r^{2}}\frac{d}{dr}\left( r^{2}\frac{d\phi }{dr}\right) =\Pi _{r}%
\frac{d\phi }{dr}+2\Pi _{r}^{2},
\end{equation}%
with the general solution given by
\begin{equation}
\phi (r)=\Pi_r c_1 \text{Ei}(\Pi_r r)-\frac{c_1 e^{\Pi_r r}}{r}-2 \Pi_r r+\frac{4}{\Pi_r r}+c_2-4 \ln (r),
\end{equation}
where $c_1$ and $c_2$ are arbitrary constants of integration, and $\text{Ei}(z)$ is the exponential function, defined as $\text{Ei}(z)=-\int_{-z}^\infty{e^{-t}dt/t}$. In the limit $\Pi_r\rightarrow 0$, we obtain the expression of the Newtonian potential $\phi (r)=-c_1/r+c_2$, which allows us to interpret $c_1$ as the mass of the gravitating object, $c_1=GM$. For small values of $\Pi_r$ we obtain
\begin{equation}\label{pot}
\phi (r)\approx -\frac{c_1}{r}+\frac{4}{\Pi_r r}+\left[c_2-4 \ln (r)\right]+\Pi_r \left[c_1 \ln (\Pi_rr)
   +(\gamma -1) c_1-2 r\right]+O\left(\Pi_r^2\right),
\end{equation}
where $\gamma$ is Euler's constant, $\gamma =0.577$. As for the expression of the Newtonian force $F_N$ we obtain
\begin{equation}\label{force}
F_N(r)=\frac{c_1 e^{\Pi_r r}}{r^2}-\frac{4}{\Pi_r r^2}-2 \Pi_r-\frac{4}{r}.
\end{equation}
Hence, the presence of the corrections to the Newtonian potential, and to the gravitational force may allow the possibility of the experimental or observational testing of the existence of the torsion of the space time. 
\subsection{Keplerian velocity and the dark matter problem.} An important physical parameter in astrophysics and astronomy is the Keplerian velocity $v_K$, which is obtained in standard Newtonian mechanics by requiring the equality of the centripetal and of the gravitational force, $v_K^2(r)/r=GM/r^2$, giving $v_K(r)=\left(GM/r\right)$, with the property $\lim_{r\rightarrow \infty}v_K(r)=0$. By equating the expression of the force as given by Eq. (\ref{force}) with the centripetal acceleration, we obtain
\begin{equation}\label{vK}
v^2_K\approx \frac{GM e^{\Pi_r r}}{r}-\frac{4}{\Pi_r r}-2 \Pi_r r-4.
\end{equation}
In the presence of torsion in the large $r$ limit the Keplerian velocity does not tend to zero, due to the presence of the exponential factor in the first term.  However, a large number of precise astrophysical investigations of the galactic rotation curves \cite{DM1,DM2,DM3,DM4} have shown that Newton’s theory of gravitational interaction, as well as standard general relativity, cannot explain the galactic dynamics of massive test particles rotating around the centers of galaxies. To describe and interpret the observational properties of the galactic rotation curves, and to also solve the missing mass problem in clusters of galaxies, the existence of a dark constituent of the Universe, called dark matter, is postulated.  Dark matter forms a static, spherically symmetric halo around the galaxies, and its interaction with baryonic matter is only gravitational. The behavior of the rotation curves of spiral galaxies \cite{DM1,DM2,DM3,DM4}  is an important problem Newtonian gravity and/or standard general relativity faces on the galactic/intergalactic scales. In spiral galaxies neutral hydrogen clouds are detected at large distances from the galactic center, beyond the presence of the luminous matter. The clouds gravitate in circular orbits with velocity $v_K(r)$, and thus in the framework of Newtonian mechanics the mass profile of the galaxy is given by
$ M(r) = rv_K^2/G$. A large number of astrophysical observations have shown that the tangential Keplerian velocities increase near the galactic center, as required by  the Newtonian theory, but after that they become nearly constant, with values of the order of $v_{K\infty}^2 \sim 200 - 300$
km/s \cite{DM1,DM2,DM3,DM4}, leading to  a mass profile $M(r) = rv_{K\infty}^2/G$. Hence, the mass of a galaxy increases linearly with $r$, even at distances where no or very little luminous matter is observed. But as shown by Eq.~(\ref{vK}), in the presence of the Semi-Symmetric torsion, the Keplerian velocity of massive test particles does not tend to zero even in the absence of baryonic matter. Thus, the presence of torsion may explain the galactic dynamics without resorting to the mysterious and not detected dark matter, which may prove to be a purely geometric effect induced by the torsion vector $\pi_\mu$. The second fundamental evidence for dark matter, after the behavior of the galactic rotation curves, comes from the virial mass discrepancy in  clusters of galaxies \cite{DM2}. A   cluster of galaxies is an astrophysical system formed of hundreds to thousands of galaxies, bounded together by their own gravitational interaction. Around 1\% of the  mass of the clusters is represented by the galaxies, the high temperature intracluster gas represents around 9\% of the cluster mass, while the dominant component is represented by dark matter, representing
90\% of the cluster mass \cite{KB}. By measuring the velocity dispersions of the galaxies one arrives to the result that the total mass of the cluster is larger than the total masses of the stars in the cluster by factors of the order of $\sim$200 - 400 \cite{KB}. Another strong evidence for the presence of dark matter is represented by the measurement of the temperature of the intracluster medium. This is due to the fact that to explain the depth of the gravitational potential of the clusters a supplementary mass component is necessary \cite{KB}. Therefore,
since clusters of galaxies are dark matter dominated objects, they represent ideal testing grounds for the properties of dark matter, or of modified gravity theories. The extra geometric terms in the field equations of the Semi-Symmetric Metric Gravity theory generate an effective contribution to the gravitational potential, as given by Eq.~(\ref{pot}). One possibility to study the effects of the Semi-Symmetric Metric gravity theory on clusters of galaxies is via the virial theorem, $2K+\Omega=0$, where $K$ is the total kinetic energy of the galaxies, and $\Omega$, the total gravitational potential energy of the cluster, can be represented as $\Omega =\Omega _B+\Omega _{G}$, where $\Omega _B$ is the baryonic matter contribution, while $\Omega _G$ denotes the contribution due to the presence of the torsion. The expressions of these terms can be obtained explicitly by using, for example, the approach developed in \cite{Vir}, which requires the use of the relativistic collisionless Boltzmann equation describing galactic motion. Hence, the study of the clusters of galaxies via the generalized virial theorem can become an efficient method in observationally testing the viability of the Semi-Symmetric Metric Gravity theory.
\subsection{Thermodynamic interpretation of the Semi-Symmetric Metric Gravity theory}
We will briefly discuss in the following the physical and thermodynamical properties of the non-conservative Semi-Symmetric Metric Gravity theory.
\subsubsection{Energy and momentum balance equation}
We will consider the matter content as consisting of a perfect fluid, whose thermodynamic properties can be described by two quantities only, the energy density $\rho$, and the thermodynamic pressure $p$, respectively. Thus, for the energy-momentum tensor of the fluid we adopt the form
\begin{equation}
T_{\mu \nu}=(\rho+p)u_\mu u_\nu-pg_{\mu \nu},
\end{equation}
where the four-velocity is normalized according to the relation $u_\mu u^\mu =1$.
To obtain the energy balance equation in Semi-Symmetric Metric Gravity we multiply both sides of
Eq.~(\ref{23}) by $u^{\mu }$. For the left hand side we obtain
\begin{eqnarray}
u_{\mu }\accentset{\circ}{\nabla} _{\nu }T^{\mu \nu } &=&u_{\mu }\accentset{\circ}{\nabla} _{\nu }\left(
\rho +p\right) u^{\mu }u^{\nu }+u_{\mu }\left( \rho +p\right)
\accentset{\circ}{\nabla} _{\nu }(u^{\mu }u^{\nu })-u_{\mu }\accentset{\circ}{\nabla} ^{\mu }p  \nonumber \\
&=&u^{\nu }\accentset{\circ}{\nabla} _{\nu }(\rho +p)+\left( \rho +p\right)
(u_{\mu }u^{\nu }\accentset{\circ}{\nabla} _{\nu }u^{\mu }+u_{\mu }u^{\mu }\accentset{\circ}{\nabla} _{\nu
}u^{\nu })-\dot{p}  \nonumber \\
&=&\dot{\rho}+\left( \rho +p\right) \accentset{\circ}{\nabla} _{\nu }u^{\nu },
\end{eqnarray}%
where we have defined $\dot{\rho}=u^{\mu }\accentset{\circ}{\nabla} _{\mu }\rho =%
\mathrm{d}\rho /\mathrm{d}s$, and we have used the relations $u_{\mu }u^{\mu }=1$, and $%
u_{\mu }u^{\nu }\accentset{\circ}{\nabla} _{\nu }u^{\mu }=0$, respectively. Therefore
the energy balance equation in the Semi-Symmetric Metric Gravity theory is obtained in the nonconservative case as
\begin{equation}\label{dotrho}
\dot{\rho}+(\rho +p)\accentset{\circ}{\nabla} _{\mu }u^{\mu }=u_{\mu}f^{\mu}.
\end{equation}
By denoting $\accentset{\circ}{\nabla} _{\mu }u^{\mu }=3H$, we can rewrite the energy-balance equation as
\begin{equation}\label{dotrho1}
\dot{\rho}+3(\rho +p)H=u_{\mu}f^{\mu}.
\end{equation}
The projection operator $h_{\thickspace \lambda}^\nu$, is defined as
\begin{equation}
 h_{\thickspace \lambda}^\nu \equiv
\delta_{\thickspace \lambda}^\nu - u^\nu u_\lambda,
\end{equation}
and it has the properties
\begin{equation}
u_\nu h_{\thickspace \lambda}^\nu = 0, h_{\thickspace %
\lambda}^\nu \accentset{\circ}{\nabla}_\mu u_\nu = \accentset{\circ}{\nabla}_\mu u_\lambda,
\end{equation}
and
\begin{equation}
h^{\nu \lambda}
\accentset{\circ}{\nabla}_\nu = \left( g^{\nu\lambda} - u^\nu u^\lambda \right) \accentset{\circ}{\nabla}_\nu =
\accentset{\circ}{\nabla}^\lambda - u^\lambda u^\nu \accentset{\circ}{\nabla}_\nu,
\end{equation}
respectively. We multiply now Eq.~(\ref{23}) with $h_{\thickspace \lambda}^\nu$,  and thus we find the momentum balance equation of the Semi-Symmetric Metric Gravity theory, which shows that the equation of motion of massive test particles is nongeodesic, and it is given by
\begin{eqnarray}\label{force0}
u^{\nu} \accentset{\circ}{\nabla}_{\nu} u^{\lambda} &=&\frac{\mathrm{d}^2x^{\lambda}}{\mathrm{d}%
	s^2}+\Gamma_{\mu \nu}^{\lambda }u^{\mu }u^{\nu} = \frac{h^{\nu \lambda} }{\rho + p}\left[f_{\nu}-
\nabla_{\nu} p \right] := Q^{\lambda}.
\end{eqnarray}
The term $Q^{\lambda}$ can be interpreted physically as an extra force acting on massive test particles in the presence of torsion. The extra force is  perpendicular to the four-velocity, $Q^{\lambda}u_{\lambda}=0$. Hence, in the presence of torsion, the dynamical motion of massive test particles is more complex than in standard general relativity.
\subsubsection{Particle creation processes in Semi-Symmetric Metric Gravity}
The non-conservation of the matter energy-momentum tensor in the Semi-Symmetric Metric Gravity, as suggested by Eq.~(\ref{23}), may be interpreted as pointing towards the existence of  particle generation processes taking place on a microscopic scale. These processes could take place locally, at the scale of the Solar System, and during the cosmological evolution. Particle creation  effects also appear in the formulation of quantum field theories in curved space-times, as shown in the pioneering studies
\cite{Parker,Parker1,Star1, Parker2}. In quantum field theory particle creation in a curved spacetime is a direct consequence of the variation in time of the gravitational field. An important result related to the finite regularized average value of the energy-momentum tensor of a quantum scalar field in anisotropic geometries, which describes both particle creation and vacuum polarization, was presented in \cite{Star1}. However, in the nonconservative version of the Semi-Symmetric Metric Gravity, particle creation is general phenomenon, which is not related to the time variability of the fields. Thus, Semi Symmetric Metric Gravity, which can consistently describe matter creation, could also help in obtaining an effective semiclassical description of quantum field processes in curved spacetimes. In the description of particle creation processes in the following in the following we use the formalism introduced in \cite{Ha14}.
\subparagraph{Thermodynamic quantities in the presence of particle creation.} Particles creation in a classical field theory is the theoretical result of the nonconservation of the matter energy-momentum tensor, whose covariant divergence does not vanish identically. Consequently, the  particle number, energy density and entropy fluxes also 
do not vanish. In the presence of particle creation the thermodynamic equilibrium equations must be extended so that they 
specifically contain in the fundamental evolution equations matter creation processes \cite{P-M,Lima,Su}. In the presence of matter creation from the gravitational field,  due to the energy transfer from geometry to particles, the particle flux
$N^{\mu} \equiv nu^{\mu}$, where $n$ is the particle number density, balance equation takes the form 
\begin{equation}\label{n}
\nabla _{\mu}N^{\mu}=\dot{n}+3Hn=n\Psi,
\end{equation}
where by $\Psi $ we have denoted the particle creation rate. Assuming that matter is in the form of dust, with the energy density $\rho =m_0n$, where $m_0$ is the particle mass, Eq.~(\ref{n}) can be reformulated in terms of density as
\begin{equation}
\dot{\rho}+3H\rho =\Psi \rho.
\end{equation}
If  $\Psi \ll H\equiv \accentset{\circ}{\nabla} _{\mu }u^{\mu }/3$, matter creation can be neglected in any physical or cosmological model. The entropy flux vector $S^{\mu}$ is defined as $S^{\mu} \equiv \tilde{s}u^{\mu} = n\sigma u^{\mu}$, where $\tilde{s}$ denotes the entropy density, and $\sigma $ denotes the entropy per particle. In the presence of particle creation the divergence of the entropy flux is given by
\begin{equation}\label{62b}
\nabla _{\mu}S^{\mu}=n\dot{\sigma}+n\sigma \Psi\geq 0.
\end{equation}
For $\Psi =0$, the total entropy is conserved, leading to an adiabatic thermodynamic process. If the entropy per particle $\sigma $ is constant, we obtain
\begin{equation}\label{con1}
\nabla _{\mu}S^{\mu}=n\sigma \Psi =\tilde{s}\Psi \geq 0.
\end{equation}
Hence if $\sigma$ is constant, the variation of the entropy determined by the gravitational particle production processes only. Since $\tilde{s}>0$, from Eq.~(\ref{con1}) we obtain the fundamental result that the matter creation rate $\Psi$ must be always positive, $\Psi \geq 0$. Consequently, matter can be produced from the curved geometry, or the gravitational fields, however, the inverse process is not allowed. 

\subparagraph{The creation pressure.} The energy-momentum tensor of matter  must be also generalized to obtain consistency with the second law of thermodynamics. This is generally done by extending the equilibrium component $T^{\mu \nu}_\text{eq}$ of the energy-momentum tensor by adding a new term $\Delta T^{\mu \nu}$, leading to \cite{Bar}
\begin{equation}\label{64}
T^{(tot)\mu \nu}=T^{\mu \nu}_\text{eq}+\Delta T^{\mu \nu},
\end{equation}
where $\Delta T^{\mu \nu}$ represents the extra contribution due to matter creation. Assuming a simple isotropic and homogeneous geometry, $\Delta T^{\mu \nu}$ can be described by a scalar quantity. Therefore, we generally obtain  \cite{Bar}
\begin{equation}
\Delta T_{\; 0}^0=0, \quad \Delta T_{\; i}^j=-P_c\delta_{\; i}^j,i,j=1,2,3,
\end{equation}
where $P_c$ represents the creation pressure, a physical quantity describing
phenomenologically the effects of matter creation via the gravitational field on macroscopic thermodynamic processes and systems. In a covariant formulation we have \cite{Bar}
\begin{equation}
\Delta T^{\mu \nu}=-P_ch^{\mu \nu}=-P_c\left(g^{\mu \nu}-u^{\mu}u^{\nu}\right).
\end{equation}
From the above relation we obtain $u_{\mu}\nabla _{\nu}\Delta T^{(tot)\mu \nu}=3HP_c$. Thus, in the presence of particle creation, the total energy balance
equation $u_{\mu}\nabla _{\nu}T^{(tot)\mu \nu}=0$, following from Eq.~\eqref{64}, can be written as 
\begin{equation}
\dot{\rho}+3H\left(\rho+P+P_c\right)=0.
\end{equation}
The Gibbs law, which is given  by \cite{Lima}
\begin{equation}
n \tilde{T} \mathrm{d} \left(\frac{\tilde{s}}{n}\right)=n\tilde{T}\mathrm{d}\sigma=\mathrm{d}\rho -\frac{\rho+p}{n}\mathrm{d}n,
\end{equation}
where $\tilde{T}$ is the thermodynamic temperature of the system, must also be satisfied by all thermodynamic quantities.
\subparagraph{Semi-Symmetric Metric Gravity and irreversible thermodynamics.} The energy balance equation~\eqref{dotrho1} of the Semi-Symmetric Metric Gravity theory can be reformulated, after some simple algebraic transformations, as
\begin{equation}\label{76}
\dot{\rho}+3H\left( \rho +P+P_{c}\right) =0,
\end{equation}%
where we have introduced the creation pressure $P_{c}$ associated to the theory, and defined as
\begin{eqnarray}
P_c=-\frac{1}{32H}u_{\mu}f^{\mu}.
\end{eqnarray}
The generalized energy balance Eq.~\eqref{76} can also be derived from the divergence of the generalized total energy momentum tensor %
$T^{\mu \nu }$ of the matter, defined as
\begin{equation}
T^{\mu \nu }=\left( \rho +P+P_{c}\right) u^{\mu }u^{\nu }-\left(
P+P_{c}\right) g^{\mu \nu }.
\end{equation}
From the Gibbs law, by assuming that matter production is an adiabatic process,
with $\dot{\sigma}=0$, we obtain
\begin{equation}
\dot{\rho}
=\left(\rho+p\right)\frac{\dot{n}}{n}
=\left(\rho+P\right)\left(\Psi-3H\right).
\end{equation}
By  using the energy balance equation we find the 
relation giving the particle production rate as a function of the creation pressure and the divergence of the matter energy-momentum tensor as 
\begin{equation}
\Psi=\frac{-3HP_c}{\rho+p}=\frac{u_{\mu}f^{\mu}}{\rho +p}.
\end{equation}
The condition $\Psi \geq 0$ imposes an important constraint on the physical parameters of Semi-Symmetric Metric Gravity theory, which can be formulated as
\begin{equation}
\frac{u_{\mu}f^{\mu}}{\rho +p}\geq 0.
\end{equation}
In terms of the creation pressure the divergence of the entropy flux vector is found as
\begin{equation}
\nabla _{\mu}S^{\mu}=\frac{-3 n \sigma H P_c}{\rho +p}=\frac{\gamma^2}{\xi^2}\frac{n\sigma }{\rho+p}u_{\mu}Q^{\mu}.
\end{equation}
\subparagraph{The temperature evolution.}
To obtain the evolution of the temperature in the presence of particle creation in a thermodynamic system we introduce the equation of state for the matter density and pressure in the general form $\rho =\rho (n, \tilde{T} )$ and $p=p(n,\tilde{T})$, respectively. Then for the variation of the energy density we obtain
\begin{equation}
\dot{\rho}=\left(\frac{\partial \rho }{\partial n} \right)_{\tilde{T}}\dot{n}+\left(%
\frac{\partial \rho }{\partial \tilde{T}} \right)_n\dot{\tilde{T}}.
\end{equation}
By using of the energy and particle balance equations we find
\begin{equation}\label{78a}
\begin{split}
-3H\left(\rho + p + P_c\right) &= \left(\frac{\partial \rho}{\partial n}\right)_{\tilde{T}} n \left(\Psi - 3H\right) \\
&\quad + \left(\frac{\partial \rho}{\partial \tilde{T}}\right)_n \dot{\tilde{T}}.
\end{split}
\end{equation}
With the help of the following thermodynamic identity \cite{Bar}
\begin{equation}\label{Termid}
\tilde{T}\left(\frac{\partial p}{\partial \tilde{T}}\right)_n=\rho+p-n\left(\frac{\partial
\rho}{\partial n}\right)_{\tilde{T}}\;,
\end{equation}
from Eq.~\eqref{78a} we obtain for the temperature evolution of a thermodynamic system  in
the presence of matter production the equation
\begin{equation}
\frac{\dot{\tilde{T}}}{\tilde{T}} = \left(\frac{\partial p}{\partial \rho}\right)_n\frac{\dot{n}}{n}=c_s^2\frac{\dot{n}}{n}=c_s^2\left(\Psi -3H\right) 
=-3Hc_s^2\left(1+\frac{P_c}{\rho +p}\right)=3Hc_s^2\left[\frac{u_{\mu}f^{\mu}}{3H(\rho+p)}-1\right],
\end{equation}
where by $c_s^2=\left(\partial p/\partial \rho \right)_n$ we have denoted the sound speed in the matter medium. If the condition $\left(\partial p/\partial \rho\right)_n=c_s^2=\mathrm{
constant}$ is satisfied, the temperature evolution in the presence of particle creation is given by
 $\tilde{T} \sim n^{c_s^2}$.
\subparagraph{The case $w=-1$.}
In our previous analysis of particle creation from the geometry, and gravitational fields, we have assumed that particles are generated in the form of baryonic matter, satisfying the condition $w=p/\rho \geq 0$. However, matter creation processes in the Semi-Symmetric Metric Gravity theory  can be generalized to the case $w<0$, corresponding to the creation of exotic forms of matter, like, for example, dark energy. In the following we will show that the thermodynamic formalism presented in the previous sections  can be applied even in the case of $w=-1$, that is, for exotic matter satisfying the equation of state $\rho +p=0$. In this case also one obtains well-defined and regular results.
As a first step in our analysis we consider the temperature evolution equation,
\begin{equation}\label{tempevol}
\frac{\dot{\tilde{T}}}{\tilde{T}}
=\left(\frac{\partial p}{\partial \rho}\right)_n\frac{\dot{n}}{n}\;,
\end{equation}
and we will prove that it is valid for $w= p/\rho = -1$. The perfect fluid energy-momentum balance equation
\begin{eqnarray}\label{61}
&&\dot{\rho}+3(\rho + p) H = u_{\mu}f^{\mu}.
\end{eqnarray}
 becomes, for $w=-1$, 
\begin{equation}
  \dot{\rho} = u_{\mu}f^{\mu}\equiv -3H P_c.
\end{equation}
 If the particle creation processes are adiabatic, with $\dot{\sigma}=0$, from the Gibbs law we find
\begin{equation}
\dot{\rho} = (\rho+P)\frac{\dot{n}}{n} = 0,
\end{equation}
and thus the above two equations give 
$\dot{\rho} = P_c = 0$.
Since $\rho = \rho \left(n, \tilde{T}\right)$, for the variation of the energy density we obtain
\begin{equation}
\dot{\rho}=\left(\frac{\partial \rho }{\partial n}\right)_{\tilde{T}} \dot{n} +\left(\frac{\partial \rho}{\partial \tilde{T}}\right)_n\dot{\tilde{T}}
= 0.
\end{equation}
From the thermodynamic identity (\ref{Termid}) with $\rho +p=0$, we find 
\begin{equation}
\tilde{T}\left(\frac{\partial P}{\partial \tilde{T}}\right)_n
=  -n\left(\frac{\partial\rho}{\partial n}\right)_{\tilde{T}}.
\end{equation}
Thus we have shown that Eq.~\eqref{tempevol} with $\left(\partial p/\partial \rho\right)_n<0$ can also be applied for $w=-1$, and for any negative values of $w$. In the case $w=-1$, Eq.~\eqref{tempevol} shows that $n\tilde{T}$ is a constant, or $\tilde{T}\sim 1/n$. Hence, if the energy density of the "dark energy" particles is very small, their  temperature must be very high. On the other hand, high density  systems composed of dark energy particle have a very low temperature. If $n\rightarrow \infty$, the dark energy particles systems have zero limiting temperature.

\section{Semi-Symmetric Metric Gravity cosmology}\label{sect2}

In the framework of Semi-Symmetric Metric Gravity, the late-time acceleration could be explained through cosmological models, which do not involve directly the cosmological constant. The accelerated expansion is entirely due to the presence of torsion, and hence has a geometric origin.
To present these models, let us assume an isotropic, homogeneous and spatially flat FLRW metric
\begin{equation}
    ds^2=-dt^2+a(t)^2 \delta_{ij} dx^i dx^j, \; \; i,j=\overline{1,3},
\end{equation}
where $i,j=\overline{1,3}$ means that $i,j$ are spatial indices, i.e. they take values in the set $\{1,2,3\}$. The matter in the universe is assumed to be composed of a perfect fluid, which appears in the Einstein equations in the form
\begin{equation}
    T_{\mu \nu}=\rho u_\mu u_\nu +p (u_\mu u_\nu+ g_{\mu \nu}),
\end{equation}
where the four velocity in a comoving frame is given by
\begin{equation}
    u_\mu=(-1,0,0,0), \; \; u^\mu=(1,0,0,0).
\end{equation}
 In accordance with the cosmological principle, as pointed out by Tsamparlis \cite{TSAMPARLIS197927}, the torsion vector reads
 \begin{equation}
     \pi^\mu=(\omega(t),0,0,0).
 \end{equation}
 The Friedmann equations of Semi-Symmetric Metric Gravity contain additional, torsion-dependent terms \cite{CSM}
 \begin{eqnarray}  \label{F1}
\hspace{-0.5cm}3H^{2}=8\pi \rho -3 \omega^2 +6H \omega=8\pi \left( \rho
+\rho _{eff}\right) =8\pi \rho _{tot},
\end{eqnarray}
\begin{eqnarray}  \label{F2}
2\dot{H}+3H^{2}=-8\pi p+4H\omega -\omega ^{2}+2\dot{\omega}=-8\pi \left(
p+p_{eff}\right)  
=-8\pi p_{tot},
\end{eqnarray}
where we have introduced the Hubble parameter $H=\frac{\dot a}{a}$, and
\begin{equation}
\rho _{eff}=\frac{1}{8 \pi} \left( 6H \omega -3 \omega^2 \right) , \; \; p_{eff}=-\frac{1}{8\pi }\left( 4H\omega -\omega ^{2}+2\dot{\omega}\right) 
\end{equation}
respectively, while $\rho _{tot}=\rho +\rho _{eff}$, and $p_{tot}=p+p_{eff}$. In this theory, as previously pointed out, the energy-momentum tensor is generally not conserved, leading to the modified continuity equation
\begin{equation}
\dot{\rho}+3H\left( \rho +p\right) +\dot{\rho}_{eff}+3H\left( \rho
_{eff}+p_{eff}\right) =0,
\end{equation}%
which can be rewritten in the equivalent form
\begin{equation}\label{85}
    \dot{\rho}+3H\left( \rho +p\right)  +\frac{3}{8\pi }\left[ \frac{d}{dt}\left(
2H\omega -\omega ^{2}\right) +2H\left( H\omega -\omega ^{2}-\dot{\omega}\right) %
\right] =0. 
\end{equation}
As an indicator of the accelerating/decelerating nature of the cosmological evolution we consider the deceleration parameter $q$, defined according to
\begin{equation}
q=\frac{d}{dt}\frac{1}{H}-1.
\end{equation}
For $p=0$, Eq.~(\ref{85}) can be reformulated as
\begin{equation}
\dot{\rho}+3H\rho =\Psi \rho,
\end{equation}
where the particle creation rate $\Psi$ is given by
\begin{equation}
\Psi=-\frac{3}{8\pi }\frac{1}{\rho }\left[ \frac{d}{dt}\left(
2H\omega -\omega ^{2}\right) +2H\left( H\omega -\omega ^{2}-\dot{\omega}\right) %
\right]. 
\end{equation}
The thermodynamic condition $\Psi \geq 0$ imposes on the cosmological quantities the constraint $\Psi \geq 0$, which gives for the cosmological quantities of the models the restriction
\begin{equation}
\omega \left(\frac{\dot{H}}{H^2}+1-\frac{\dot{\omega}}{H^2}-\frac{\omega}{H}\right)\leq 0.
\end{equation}
In terms of the deceleration parameter we obtain the restriction
\begin{equation}
\omega \left(q+\frac{\dot{\omega}}{H^2}+\frac{\omega }{H}\right)\geq 0.
\end{equation}
As for the creation pressure, we obtain the general expression
\begin{equation}
P_c=-\frac{\omega H}{4\pi}\left(q+\frac{\dot{\omega}}{H^2}+\frac{\omega }{H}\right).
\end{equation}
\subsection{Specific cosmological models}
In the following we will consider two simple cosmological models in the Semi-Symmetric Metric Gravity theory, obtained by imposing an equation of state on the effective dark geometric pressure. With this assumption the set of the cosmological field equations can be closed, and their solution can be obtained by using numerical methods. However, as the purpose of the cosmological models is to compare with observational data, let us rewrite the Friedmann equations in two steps, first in dimensionless variables, then in redshift variables. We introduce dimensionless parameters as
\begin{equation}
H=H_{0}h,\tau =H_{0}t,\omega =H_{0}\Omega ,\rho =\frac{3H_{0}^{2}}{8\pi }r,p=%
\frac{3H_{0}^{2}}{8\pi }P.
\end{equation}
With the help of these, we can rewrite the Friedmann equations in the form
\begin{equation}
h^{2}=r- \Omega^2+ 2 h \Omega,
\end{equation}
\begin{equation}
2\frac{dh}{d\tau }+3h^{2}=-3P+4h\Omega -\Omega ^{2}+2\frac{d\Omega }{d\tau},
\end{equation}%
while the effective energy density and pressure become
\begin{equation}
r_{eff}=2h\Omega -\Omega ^{2},
\end{equation}
\begin{equation}
P_{eff}=-\frac{1}{3}\left( 4h\Omega -\Omega ^{2}+2\frac{d\Omega }{d\tau }%
\right) ,
\end{equation}
where $\rho _{eff}=\left( 3H_{0}^{2}/8\pi \right) r_{eff}$, and $%
p_{eff}=\left( 3H_{0}^{2}/8\pi \right) P_{eff}$. The redshift parametrization is defined by $1+z=\frac{1}{a}$, which directly implies
\begin{equation}
    \frac{d}{d \tau}=-(1+z)h(z) \frac{d}{dz}.
\end{equation}
Hence, in the redshift parametrization, the Friedmann equations are given by
\begin{equation}
h^{2}(z)=r(z)+2h(z)\Omega (z)-\Omega ^{2}(z),  \label{Fz1}
\end{equation}%
\begin{equation}  \label{Fz2}
-2(1+z)h(z)\frac{dh(z)}{dz}+3h^{2}(z)=-3P(z)+4h(z)\Omega (z)  
-\Omega ^{2}(z)-2(1+z)h(z)\frac{d\Omega }{dz}.
\end{equation}
\subsubsection{Linear cosmological model}
So far, the torsion field $\pi$ is not provided with dynamics. To give dynamics to the field, we assume two equations of state, one for the ordinary matter, and one for the effective components. As a first cosmological model, we assume $p=0$, that is, ordinary matter is a pressureless dust. For the effective components, an equation of state of the form
\begin{equation}
P_{\text{eff}}(z) = -\sigma_0 \rho_{\text{eff}}(z).
\end{equation}
is proposed. Given these, the system of differential equations governing the evolution of the Universe in redshift representation, is given by
\begin{equation}\label{linear1}
\frac{dh(z)}{dz} = \frac{3h^2(z) - 3\sigma_0 (2h(z)\Omega(z) - \Omega^2(z))}{2(1 + z)h(z)}
\end{equation}
\begin{equation}\label{linear2}
\frac{d\Omega(z)}{dz} = \frac{-2(3\sigma_0 - 2)h(z)\Omega(z) - (1 - 3\sigma_0)\Omega^2(z)}{2(1 + z)h(z)}
\end{equation}
The system of equations has to be solved with initial conditions $h(0) = 1$, $\Omega(0)=\Omega_0$. Then, from the closure relation, the matter density can be obtained as
\begin{equation}
    r(z)=h^2(z)-2h(z) \Omega(z) + \Omega^2(z)=\left(h(z)-\Omega (z)\right)^2,
\end{equation}
or
\begin{equation}
\Omega (z)=h(z)\mp \sqrt{r(z)}.
\end{equation}
Estimating the matter density at the present time gives
\begin{equation}
r(0)=\left(1-\Omega _0\right)^2,
\end{equation}
or
\begin{equation}
\Omega (0)=1\mp \sqrt{r(0)}.
\end{equation}
Hence, the present day value of the torsion vector is fully determined by the present day value of the matter density.
\subsubsection{Polytropic cosmological model}
As a second cosmological model, we keep the assumption of $p=0$, but consider a different equation of state for the effective components, namely a polytropic EoS
\begin{equation}
    P_{eff}=K \left(\rho_{eff}\right)^{{\left(1+\frac{1}{n} \right)}}.
\end{equation}
where $K$ is a constant, and $n$ is the polytropic index. It is important to note that for matter we still consider a pressureless dust, and we impose the polytropic equation of state only for the effective components. In this model, the evolution equations in the redshift representation take the form
\begin{equation}\label{polytropic1}
\frac{d\Omega(z)}{dz} = \frac{1}{2(1 + z)h(z)} \left( 4h(z)\Omega(z) - \Omega^2(z) + 3K \left( 2h(z)\Omega(z) - \Omega^2(z) \right)^{\left(1+\frac{1}{n} \right)} \right)
\end{equation}
\begin{equation}\label{polytropic2}
\frac{dh(z)}{dz} = \frac{3h^2(z) - 4h(z)\Omega(z) + \Omega^2(z) + 2(1 + z)h(z) \frac{d\Omega(z)}{dz}}{2(1 + z)h(z)}
\end{equation}
They have to be integrated with the initial conditions $h(0) = 1$, $\Omega(0)=\Omega_0$. In this particular case, the polytropic index is chosen to be $n = \frac{3}{5}$. The first Friedmann equation yields again the expression for the matter density
\begin{equation}
    r(z)=h^2(z)-2h(z) \Omega(z) + \Omega^2(z)=\left[h(z)-\Omega (z)\right]^2.
\end{equation}
\subsection{Observational constraints}

In this study, we use a Markov Chain Monte Carlo (MCMC) approach to estimate the optimal parameters of the proposed cosmological models. The Hubble parameter is derived from a system of differential equations in both cases: in the linear case, this system is explicitly \eqref{linear1}-\eqref{linear2}, while in the polytropic case, the system is given by \eqref{polytropic1}-\eqref{polytropic2}. The numerical solutions are obtained using the \texttt{solve\_ivp} function from SciPy \cite{ODE,SciPy}. We apply the Radau method \cite{Radau}, a fifth-order implicit Runge-Kutta technique, particularly suited for stiff differential equations that frequently occur in cosmological modeling due to the wide range of scales and rapid variations across different epochs. The Radau method ensures numerical stability, especially over large redshift intervals, where the equations exhibit stiff behavior. To balance computational efficiency and accuracy, we set the relative tolerance to \(10^{-3}\) and absolute tolerance to \(10^{-6}\). These tolerance levels allow stable and accurate integration across the redshift range \(z \in [0, 3]\), guaranteeing that the solution captures both large and small variations in the variables. Having obtained the numerical solutions, in terms of the model parameters, we use a Markov Chain Monte Carlo (MCMC) algorithm to estimate the best fit values of the parameters, with respect to several data sets \cite{Bayes}. The MCMC algorithm explores the parameter space by sampling from the posterior distribution
\begin{equation}
P(\theta|D) = \frac{L(D|\theta) P(\theta)}{P(D)},
\end{equation}
where \(P(\theta|D)\) is the posterior probability of the parameters \(\theta\) given the data \(D\), \(L(D|\theta)\) is the likelihood of the data given the parameters, \(P(\theta)\) represents the prior distribution, and \(P(D)\) is the evidence, acting as a normalization constant. It is important to note that the MCMC method not only identifies the most probable or optimal parameters, but also takes into account the uncertainties in the model (this is not the case here) and in the data.
\subsubsection{Data description}
To find the optimal parameters, the following data sets were used: a set of 31 Cosmic Chronometers (CC) observations, Type Ia Supernovae (SNe Ia) data without SHOES calibration, Baryon Acoustic Oscillation (BAO) measurements from observations of galaxies, quasars, and Lyman-\(\alpha\) tracers. This includes data from the first year of the Dark Energy Spectroscopic Instrument (DESI), and the completed SDSS-IV extended Baryon Oscillation Spectroscopic Survey. Additionally, the Hubble constant, as estimated by the SH0ES team, is \( H_0 = 73.04 \pm 1.04 \, \text{km} \, \text{s}^{-1} \, \text{Mpc}^{-1} \). In the following, we briefly describe each data set, together with the corresponding likelihoods used in our analysis.
\subparagraph{Cosmic chronometers (CC).}
The 31 Cosmic Chronometer (CC) measurements cover a redshift range of approximately \(z \lesssim 2\). These measurements were obtained using the differential age method \cite{differentialage}, which involves passively evolving galaxies observed at small redshift intervals. This technique allows for the direct calculation of the Hubble parameter by measuring the rate of change of redshift with respect to cosmic time, \(\Delta z / \Delta t\). By focusing on galaxies that are not actively forming stars, this method provides a model-independent estimate of the Universe's expansion rate at different epochs. In this study, we incorporate 31 data points from various independent sources, as outlined in table 1 of \cite{cosmic chronometers}. For our MCMC analysis, we evaluate the goodness-of-fit using the \(\chi^2_{CC}\) statistic
\begin{equation}
\chi^2_{CC}(\theta) = \Delta H^T(z) \mathbf{C}^{-1} \Delta H(z),
\end{equation}
where \( \Delta H(z) \) is the vector of residuals, defined as \( \Delta H(z) = \mathbf{H_{\text{model}}}(\theta) - \mathbf{H_{\text{obs}}} \). Here, \( \mathbf{H_{\text{model}}}(\theta) \) is the vector of theoretical Hubble parameter values at redshifts \( z_i \) for the model parameters \(\theta\), and \( \mathbf{H_{\text{obs}}} \) is the vector of observed Hubble parameter values. \( \mathbf{C} \) is the covariance matrix of the observational data, with diagonal elements representing the variances \( \sigma_H^2(z_i) \) of the observed Hubble parameters. Since the data points are assumed to be uncorrelated, the covariance matrix is diagonal. \( \mathbf{C}^{-1} \) is the inverse of the covariance matrix, used to account for the measurement uncertainties.
\subparagraph{Type Ia supernova (SNe Ia).}
The Pantheon+ dataset includes light curves for 1701 Type Ia Supernovae (SNe Ia) from 1550 distinct supernovae, spanning a redshift range of \(0 \leq z \leq 2.3\) \cite{Pantheon}. The observable quantity for SNe Ia is the apparent magnitude
\begin{equation}
m(z) = 5 \log_{10} \left( \frac{d_L (z)}{\text{Mpc}} \right) + \mathcal{M} + 25,
\end{equation}
where \(\mathcal{M}\) denotes the absolute magnitude of SNe Ia. The luminosity distance \(d_L\) in a flat Friedmann-Lemaître-Robertson-Walker (FLRW) Universe is given by \cite{Pantheon}
\begin{equation}
d_L (z) = c(1 + z) \int_{0}^{z} \frac{dz'}{H(z')},
\end{equation}
with \(c\) representing the speed of light in km/s. For SNe Ia, the \(\chi^2\) statistic is computed using
\begin{equation}
\chi^2_{\text{SNe Ia}} = \Delta \mathbf{D}^T \mathbf{C}^{-1}_{\text{total}} \Delta \mathbf{D},
\end{equation}
where \(\mathbf{C}_{\text{total}} = \mathbf{C}_{\text{stat}} + \mathbf{C}_{\text{sys}}\) combines the statistical and systematic covariance matrices, and \(\Delta \mathbf{D}\) is the vector of 1701 SNe Ia distance moduli, calculated as:
\begin{equation}
\Delta \mathbf{D} = \mu(z_i) - \mu_{\text{model}} (z_i, \theta).
\end{equation}
Here, \(\mu(z_i) = m(z_i) - \mathcal{M}\) is the distance modulus of SNe Ia.
\subparagraph{Baryon Acoustic Oscillations (BAO).}
The BAO scale is determined by the sound horizon at the drag epoch \( z_d \), which marks the decoupling of photons and baryons. This scale is defined by the integral
\begin{equation}
r_d = \int_{z_d}^{\infty} \frac{c_s(z)}{E(z)} \, dz .
\end{equation}
In this context, the speed of sound in the baryon-photon fluid, \( c_s \), is approximated by:$ c_s \approx c \left( 3 + \frac{9 \rho_b}{4 \rho_\gamma} \right)^{-0.5},$
where \( \rho_b(z) \) and \( \rho_\gamma(z) \) are the baryon and photon densities, respectively \cite{photon}. The function \( E(z) \) is defined as the product of \( H(z) \) and the present-day Hubble parameter \( H_0 \), incorporating cosmological model parameters. The drag epoch, when baryons decouple from photons, occurs at approximately \( z_d \approx 1060 \). In a flat \(\Lambda\)CDM cosmology, \cite{Planck} estimates the sound horizon at drag as \( r_d = 147.09 \pm 0.26 \, \text{Mpc} \). In \cite{Verde} the value \( r_d = 143.9 \pm 3.1 \, \text{Mpc} \) is obtained. Additionally, \cite{Lemos} employs binning and Gaussian methods on 2D BAO and supernova data to estimate the absolute BAO scale, finding ranges of \( 141.45 \, \text{Mpc} \leq r_d \leq 159.44 \, \text{Mpc} \) (binning) and \( 143.35 \, \text{Mpc} \leq r_d \leq 161.59 \, \text{Mpc} \) (Gaussian). Furthermore, independent of CMB data, \cite{Nunes} finds \( r_d = 144^{+5.3}_{-5.5} \, \text{Mpc} \) (from \( \theta_{\text{BAO}} + \text{BBN} + \text{HoLiCOW} \)), while \cite{Pogosian} reports \( r_d = 143.7 \pm 2.7 \, \text{Mpc} \). In the present analysis, we remove the \( r_d \) prior from the CMB Planck satellite data, and set \( r_d \) as a free parameter. This approach provides the advantage of model-independence by avoiding specific assumptions about the early Universe and recombination processes \cite{Jedamzik,Pogosian2,Lin,Vagnozzi}. We then determine the transverse distance \( D_H(z) \) for each model, defined as follows
\begin{equation}
D_H(z) = \frac{c}{H_{0}h(z)},
\end{equation}
where \(c\) is the speed of light in  vacuum, \(H_0\) is the present-day Hubble constant, and \(h(z)\) is the numerical solution of the differential equation obtained using the initial conditions. We also calculate the comoving angular diameter distance \(D_M(z)\), which depends on the expansion history and curvature, as
\begin{equation}
D_M(z) = \frac{c}{H_0} S_k \left(\frac{D_C(z)}{c/H_0}\right).
\end{equation}

Here, the line-of-sight comoving distance \( D_C(z) \) is given by
\begin{equation}
D_C(z) = \frac{c}{H_0} \int_{0}^{z} \frac{dz'}{h(z')}.
\end{equation}

The function \( S_k(x) \) is defined as
\begin{equation}
S_k(x) =
\begin{cases} 
\frac{\sin(\sqrt{-\Omega_k} \, x)}{\sqrt{-\Omega_k}} & \text{if } \Omega_k < 0, \\
x & \text{if } \Omega_k = 0, \\
\frac{\sinh(\sqrt{\Omega_k} \, x)}{\sqrt{\Omega_k}} & \text{if } \Omega_k > 0.
\end{cases}
\end{equation}
For a flat Universe (as considered by us), the curvature parameter vanishes, i.e. \(\Omega_k = 0\), and the function \( S_k(x) \) simplifies to $S_k(x) = x.$ Thus, the formula for the comoving angular diameter distance \(D_M(z)\) in a flat Universe takes the simplified form
\begin{equation}
D_M(z) = \frac{c}{H_0} \cdot \frac{D_C(z)}{c/H_0} = D_C(z) = \frac{c}{H_0} \int_0^z \frac{dz'}{h(z')},
\end{equation}
The value of \( r_d \) can be determined by constraining the ratios \( \frac{D_M(z)}{r_d} \) and \( \frac{D_H(z)}{r_d} \). We also consider the spherically averaged distance \( D_V(z) \), defined as
\begin{equation}
D_V(z) \equiv \left[ z D_M^2(z) D_H(z) \right]^{1/3}.
\end{equation}
Additionally, we constrain the quantity \( \frac{D_V(z)}{r_d} \). In this expression, the exponent \( \frac{1}{3} \) accounts for the radial dimensions, while the factor of \( z \) ensures conventional normalization. To compute the $\chi^2_{BAO}$ function for the Baryon Acoustic Oscillation (BAO) data, we use the following equations
\begin{equation}
\chi^2_{D_X / r_d} = \Delta D_X^T \cdot \mathbf{C}^{-1}_{D_X} \cdot \Delta D_X,
\end{equation}
where \( \Delta D_X = D_{X / r_d, \text{Model}} - D_{X / r_d, \text{Data}} \) for \( X = H, M, V \), and \( \mathbf{C}_{D_X}^{-1} \) is the inverse covariance matrix corresponding to each \( X \). The inverse covariance matrix, \( \mathbf{C}_{D_X}^{-1} \), is computed by inverting the covariance matrix \( \mathbf{C}_{D_X} \), which is typically constructed by incorporating the observational uncertainties \( \sigma_{D_X} \) along the diagonal, i.e., \( \mathbf{C}_{D_X} = \text{diag}(\sigma_{D_X}^2) \). The BAO chi-square is defined as
\begin{equation}
\chi^2_{\text{BAO}} = \chi^2_{D_H / r_d} + \chi^2_{D_V / r_d} + \chi^2_{D_M / r_d}.
\end{equation}
In our analysis, we incorporate the latest Baryon Acoustic Oscillation (BAO) data from the Dark Energy Spectroscopic Instrument (DESI) Year 1 observations, as detailed in Table 1 of \cite{Desi}. Additionally, we include data from the completed SDSS-IV extended Baryon Oscillation Spectroscopic Survey, presented in Section 4 of Table 3 in \cite{SDSS}.
\subparagraph{Hubble Constant Measurements}
We also incorporate the Hubble constant value (\(H_0\)) as estimated by the SHOES collaboration \cite{SHOEScollaboration}. The SHOES collaboration reports a Hubble constant value of $H_0^{R22} = 73.04 \pm 1.04 \, \mathrm{km} \, \mathrm{s}^{-1} \, \mathrm{Mpc}^{-1}$ \cite{R22}. Using Type Ia supernovae as standard candles, the SHOES collaboration infers distances to galaxies, providing a robust estimate of the Hubble constant. We employ the R22 prior to demonstrate the tension between our model and the observed value. To incorporate the R22 prior, we define the \(\chi^2_{R22}\) statistic as 
\[
\chi^2_{R22} = -0.5 \left( \frac{H_0^{R22} - H_0^{\text{Model}}}{\sigma_{R22}} \right)^2
\]
Here, \(H_0^{R22} = 73.04\), \(H_0^{\text{Model}}\) is the theoretical Hubble constant value computed for each model using a numerical analysis, and \(\sigma_{R22} = 1.04\) is the error associated with \(H_0^{R22}\). To incorporate all the information from each datasets, we define the total chi-squared statistic, \( \chi^2_{\text{tot}} \), as the sum of the individual chi-squared contributions from each dataset: 
\[
\chi^2_{\text{tot}} = \chi^2_{\text{CC}} + \chi^2_{\text{SNe Ia}} + \chi^2_{\text{BAO}} + \chi^2_{\text{R22}}.
\]  
When we consider the combination of CC + SNe Ia + BAO, we label it as (JOINT). When the R22 prior is included, we refer to it as (JOINT + R22). The MCMC analysis is performed using the \texttt{emcee} library \cite{emcee}. For visualizing and plotting the results, we use the publicly available \texttt{GetDist} package \cite{GetDist}, which offers a comprehensive toolkit for generating 1D and 2D posterior distribution plots.
\begin{table*}
\centering
\begin{tabular}{|l|c|c|c|c|}
    \hline
    \textbf{Cosmological Models} & \textbf{Parameter} & \textbf{Prior} & \textbf{JOINT} & \textbf{JOINT + R22} \\
    \hline\hline
    \multirow{5.5}{*}{\textbf{$\Lambda$CDM Model}} & $H_{0}$ & $[50.,100.]$ & $68.0{\pm 1.6}$ & $71.48{\pm 0.87}$   \\[1ex]
    & $\Omega_{m0}$ & $[0,1.]$   & $0.322{\pm 0.011}$ & $0.296{\pm 0.011}$ \\[1ex]
    & $\mathcal{M}$ & $[-20,-18]$ & $-19.415{\pm 0.051}$ & $-19.306{\pm 0.027}$ \\[1ex]
    & $r_{d}$ & $[100,300]$ & $146.4{\pm 3.4}$ & $139.2{\pm 1.8}$ \\[1ex]
    \hline
    \multirow{5.5}{*}{\textbf{Linear Model}} & $H_{0}$ & $[50.,100.]$ & $66.9{\pm 1.6}$  & $71.28{\pm 0.89}$ \\[1ex]
    & $\Omega_0$ & $[0.,1.]$   & $0.3914{\pm 0.0095}$ & $0.3955{\pm 0.0091}$ \\[1ex]
    & $\sigma_0$ & $[0.,2.]$   & $0.950{\pm 0.047}$ & $0.979{\pm 0.046}$ \\[1ex]
    & $\mathcal{M}$ & $[-20,-18]$ & $-19.302{\pm 0.027}$ &  $-19.435{\pm 0.051}$  \\[1ex]
    & $r_{d}$ & $[100,300]$ & $146.8{\pm 3.4}$ & $138.8{\pm 1.9}$ \\[1ex]
    \hline
    \multirow{5.5}{*}{\textbf{Polytropic Model}} & $H_{0}$ & $[50.,100.]$ & $66.8{\pm 1.6}$ & $71.25{\pm 0.88}$ \\[1ex]
    & $\Omega_{0}$ & $[0.,1.]$   & $0.3808{\pm 0.0083}$ & $0.3867{\pm 0.0080}$ \\[1ex]
    & $K$ & $[-3,-1.]$   & $-1.91_{-0.17}^{+ 0.15}$ & $-1.97_{-0.14}^{+ 0.18}$  \\[1ex]
    & $\mathcal{M}$ & $[-20,-18]$ & $-19.432{\pm 0.051}$ &  $-19.296{\pm 0.027}$  \\[1ex]
    & $r_{d}$ & $[100,300]$  &  $146.8{\pm 3.4}$ & $138.6{\pm 1.9}$ \\[1ex]
    \hline
\end{tabular}
\caption{Summary of the optimal values for the cosmological parameters within the framework of Semi-Symmetric Metric Gravity.}
\label{tab_2}
\end{table*}
\begin{figure*}
\centering
\includegraphics[width=10.5 cm]{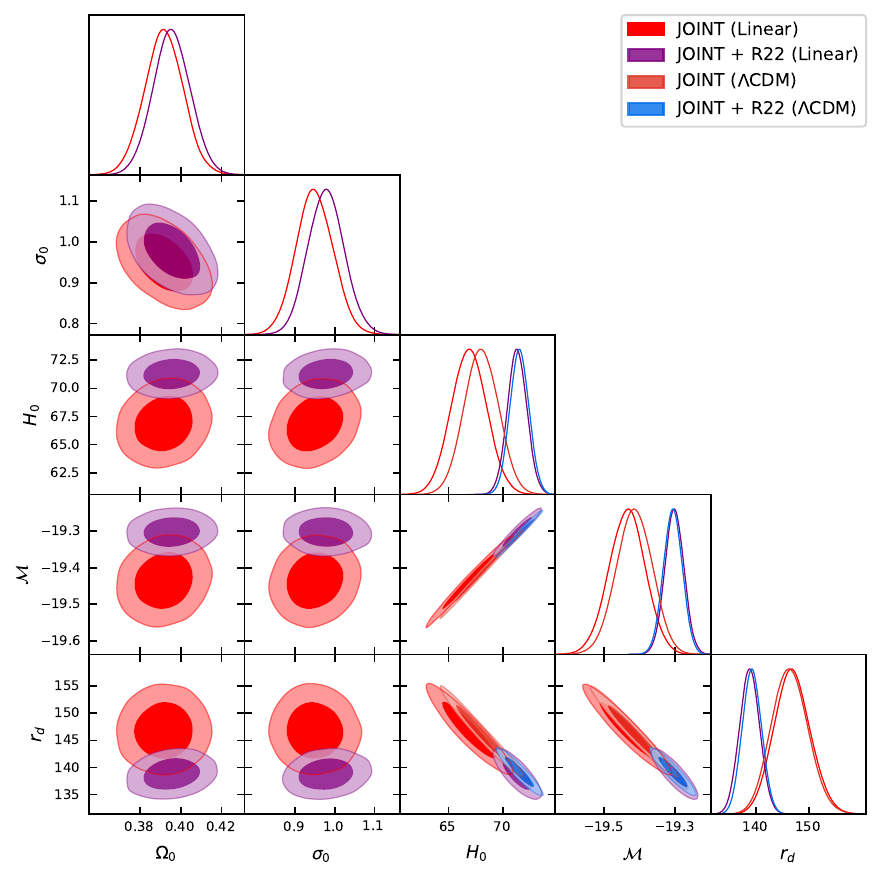}
\caption{The confidence contours at the 1\(\sigma\) and 2\(\sigma\) levels based on constraints for the linear model within the Semi-Symmetric Metric Gravity framework.}\label{fig_1}
\end{figure*}
\begin{figure*}
\centering
\includegraphics[width=10.5 cm]{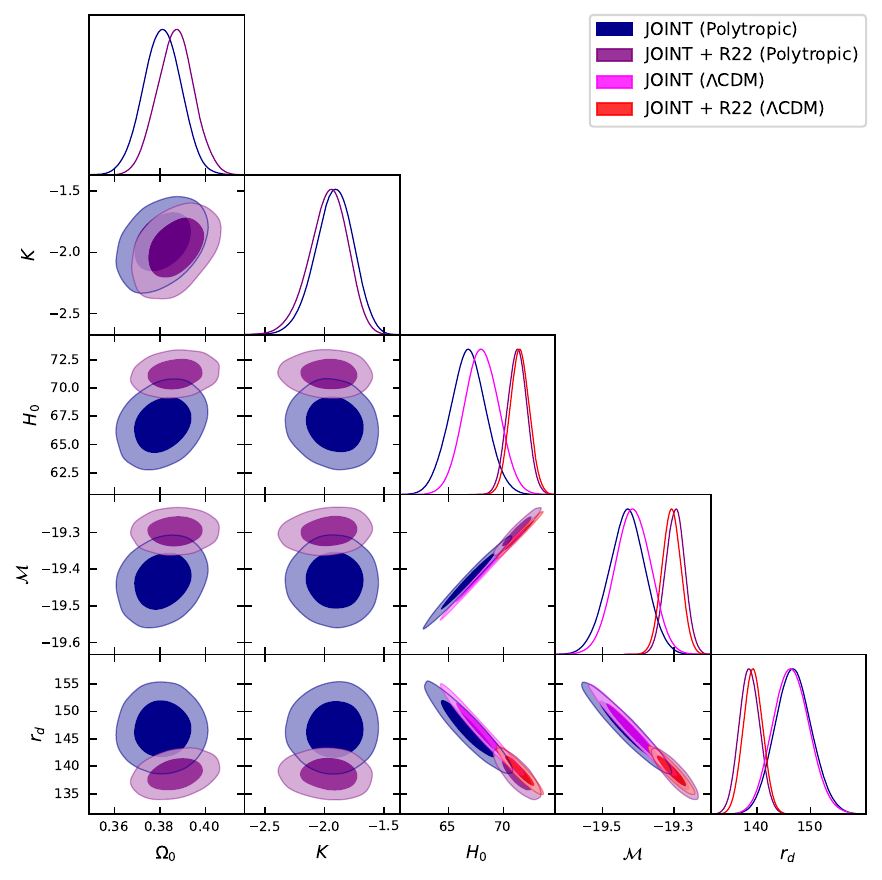}
\caption{The confidence contours at the 1\(\sigma\) and 2\(\sigma\) levels for the polytropic model within the Semi-Symmetric Metric gravity framework.}\label{fig_2}
\end{figure*}
\subsection{Analysis of the cosmological models using the MCMC-Inferred optimal parameters}
Having obtained the optimal parameter values from the MCMC algorithm, we compare our cosmological models both with the observational data (in the figures only the 31 CC are considered) and the standard $\Lambda$CDM model. We do this in several steps, comparing both the Hubble parameters, distance moduli, and several key cosmographic quantities.
\subsubsection{Comparative evolution of the Hubble parameter $H(z)$} 
For the \( \Lambda \)CDM model, we use the expression 
\begin{equation}
H_{\Lambda \text{CDM}}(z, \Omega_m, H_0) = H_0 \sqrt{\Omega_m (1 + z)^3 + (1 - \Omega_m)},
\end{equation} where \( \Omega_{m0} = 0.322 \) and the present-day Hubble constant \( H_0 = 68.0 \) km/s/Mpc, as derived from the MCMC analysis using the joint data. In our differential equation system analysis, we obtain the model parameter \( h(z) \) by numerically integrating the system of differential equations using appropriate initial conditions. Once the solution \( h(z) \) is obtained, we interpolate it using `interp1d` to determine the values of \( h(z) \) at specific redshift points corresponding to the observational data. Finally, we scale the interpolated solution \( h(z) \) by the present-day Hubble constant \( H_0 \) to compute the model Hubble parameter \( H_{\text{model}}(z) \), which is then compared with the observational data.
\begin{figure}[H]
\centering
\includegraphics[width=8.5 cm]{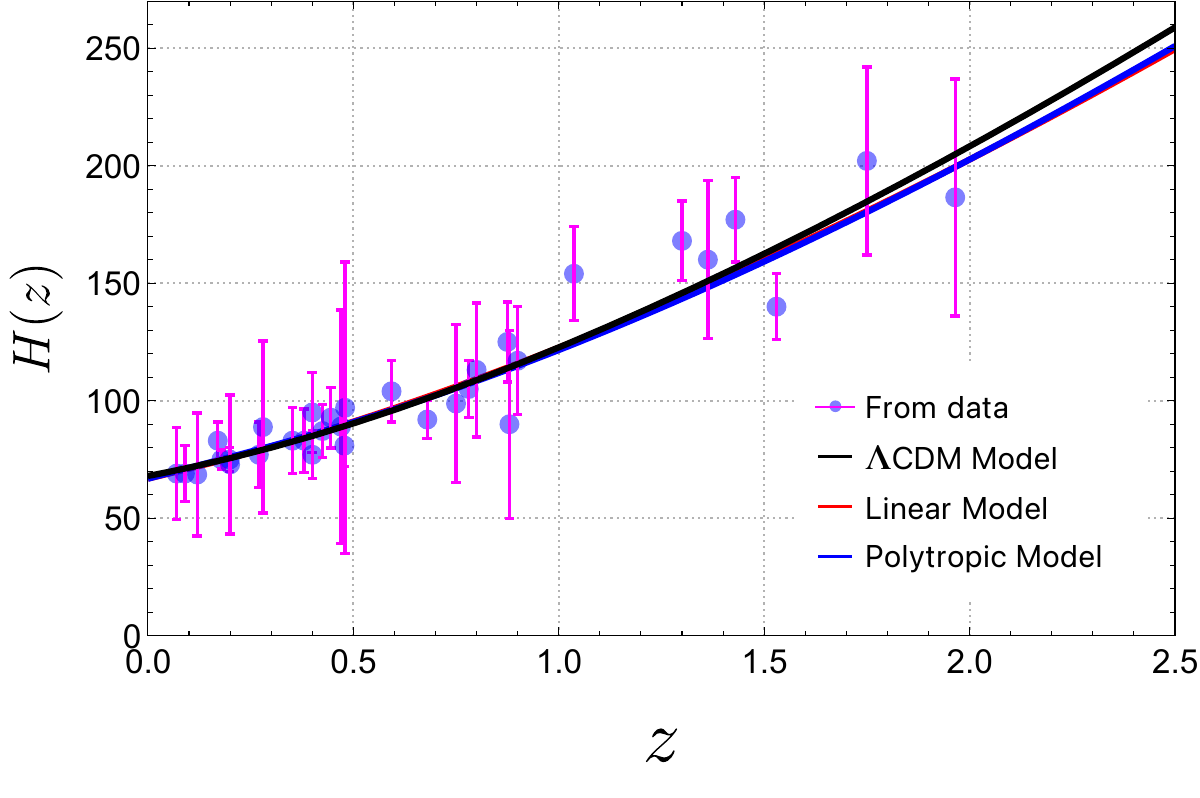}
\caption{Evolution of the Hubble parameter \(H(z)\) as a function of redshift \(z\), using the joint data best-fit values, and compared with the CC dataset.}\label{fig_3}
\end{figure}
\subsubsection{Comparative evolution of the Hubble Difference $\Delta H(z)$} 
After obtaining the Hubble parameter \( H(z) \) from our model, we compute a quantity called the Hubble difference, denoted as \( \Delta H(z) \). This is given by the formula: $\Delta H(z) = H_{\text{model}}(z) - H_{\Lambda \text{CDM}}(z)$ where \( H_{\text{model}}(z) \) is the Hubble parameter predicted by our model and \( H_{\Lambda \text{CDM}}(z) \) is the Hubble parameter predicted by the standard \( \Lambda \)CDM model. To calculate \( \Delta H(z) \) for the \( \Lambda \)CDM model itself, we need to take the difference between the observed Hubble parameter \( H(z) \) and \( H_{\Lambda \text{CDM}}(z) \) at the corresponding redshift points. This provides a measure of how much the model's predictions deviate from the standard \( \Lambda \)CDM predictions and from the observed data.
\begin{figure}
\centering
\includegraphics[width=8.5 cm]{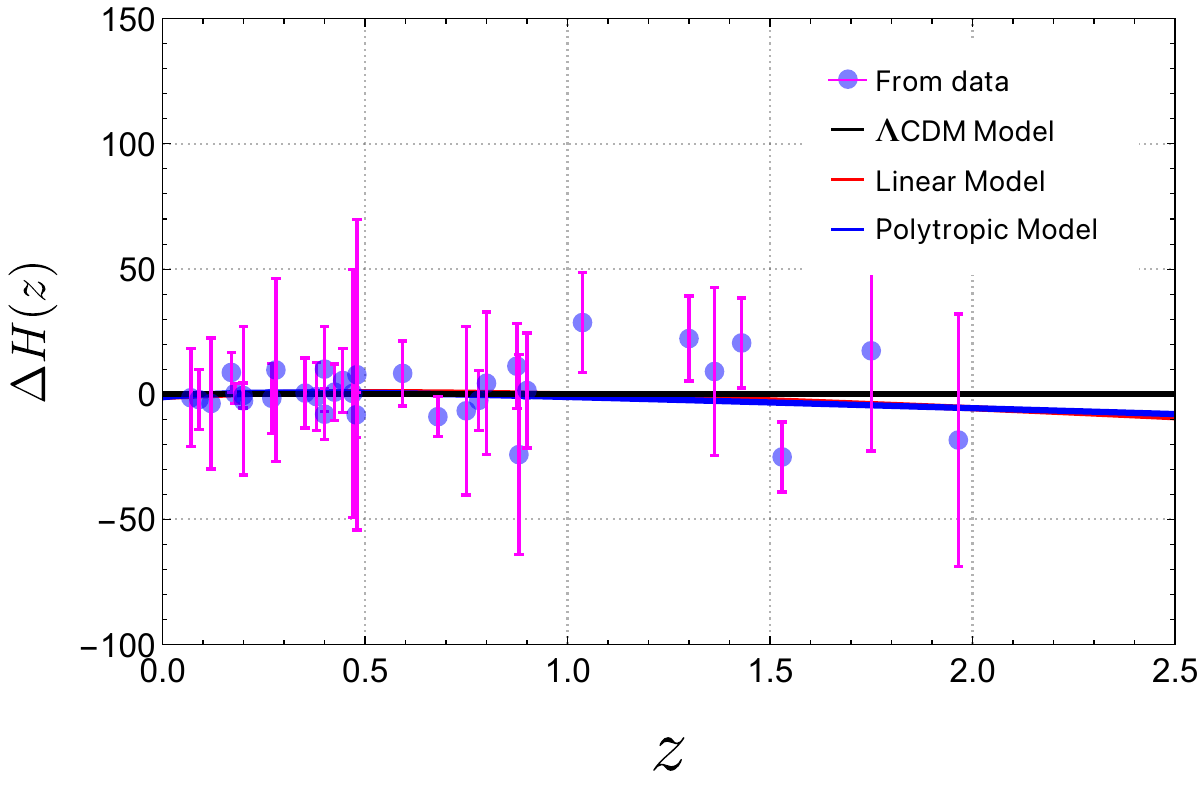}
\caption{Evolution of the Hubble difference $\Delta H(z)$ as a function of redshift \(z\), using the joint data best-fit values, and compared with the CC dataset.}\label{fig_4}
\end{figure}
\subsubsection{Comparative evolution of the distance modulus $\mu(z)$} 
The calculation of \( \mu(z) \) involves several steps: First, we calculate the comoving distance as a function of redshift \( z \). The comoving distance \( D_C(z) \) is given by the integral:
$D_C(z) = \int_0^z \frac{c}{H(z')} \, dz'$, where \( c \) is the speed of light, and \( H(z') \) is the Hubble parameter at a given redshift \( z' \). This integral is evaluated from the present time (\( z = 0 \)) to the redshift \( z \) of the object. Next, we compute the luminosity distance, \( D_L(z) \), which is related to the comoving distance by: $D_L(z) = (1 + z) D_C(z)$. The factor \( (1 + z) \) accounts for the redshift of the light as it travels from the source to the observer, with \( D_C(z) \) representing the comoving distance to the object. Finally, the distance modulus \( \mu(z) \), which relates the luminosity distance to the observed magnitude, is computed using the formula: $\mu(z) = 5 \log_{10}(D_L(z)) + 25$ Here, \( D_L(z) \) is expressed in megaparsecs (Mpc). Using the best-fit values obtained from the MCMC algorithm, we calibrated \(\mu_{\text{\(\Lambda\)CDM}}(z)\) and \(\mu_{\text{Model}}(z)\), and plotted them against the dataset of 1701 Type Ia supernovae (SNe Ia).
\begin{figure}[H]
\centering
\includegraphics[width=8.5 cm]{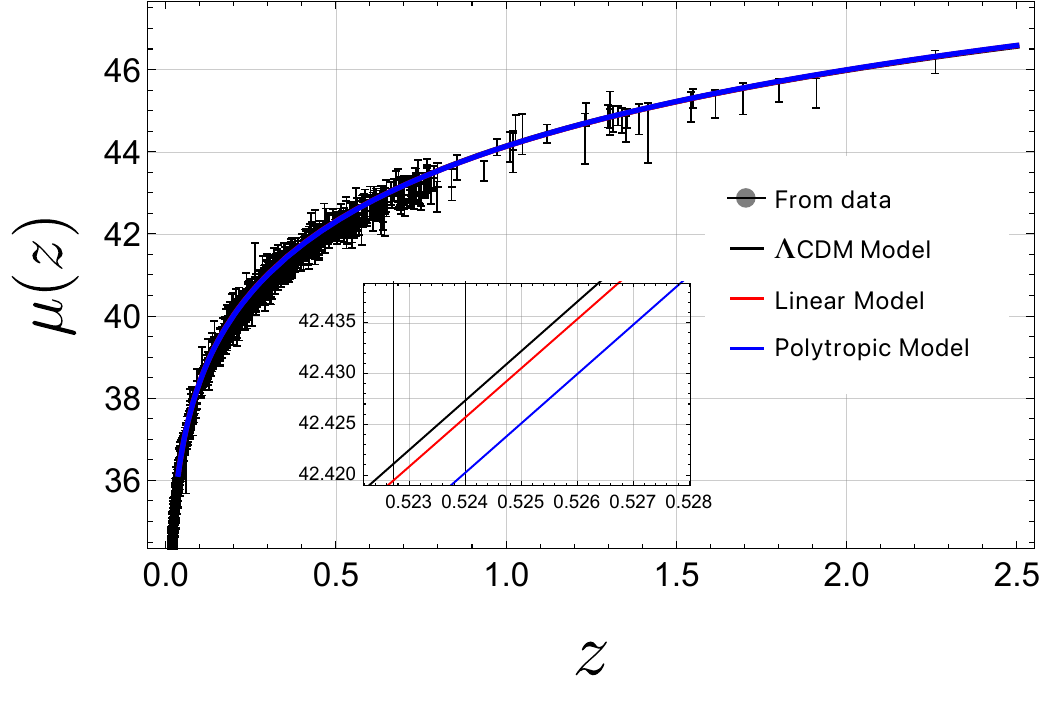}
\caption{Evolution of the distance modulus \( \mu(z) \) as a function of redshift \(z\), using the joint data best-fit values, and compared with the SNe Ia dataset.}\label{fig_5}
\end{figure}
\subsection{Cosmographic analysis}\label{sec6}
As there are many modified gravitational theories, which  could reproduce the observational data of the Hubble parameter, cosmography \cite{Cosmographic1,Cosmographic2,Cosmographic3,Cosmographic4} is used to differentiate between these models Even if a model fits the Hubble observational data, the derived quantities, usually expressed in terms of series expansions of the Hubble parameter, could differentiate between the models. Such derived quantities include the deceleration, jerk and snap parameters. These go beyond measuring the expansion rate, and offer insight into the dynamics and variations in the Universe's evolution over time. In the following, we briefly describe these derived quantities.
\subsubsection{Deceleration parameter \( q(z) \)}
The deceleration parameter \( q(z) \) is essentially used to decide whether the expansion of the Universe is accelerating or decelerating at a particular redshift. When \( q(z) \) is negative, it indicates that the Universe is accelerating, while a positive value means the expansion is slowing down. Formally, it is defined as
\[
q(z) = -\frac{1}{H^2(z)}\frac{dH(z)}{dz} - 1,
\]
where the term \( \frac{dH(z)}{dz} \) represents the rate of change of the Hubble parameter with respect to redshift. Since the sign of $q(z)$ describes the accelerating/decelerating phases of the cosmic evolution, a change of sign in $q(z)$ signals a transition between two phases. In the decelerating phase, matter dominates, while the accelerating phase is usually driven by dark energy (or the cosmological constant), but in our models, this is entirely due to the presence of torsion.
\begin{figure}[H]
\centering
\includegraphics[width=8.5 cm]{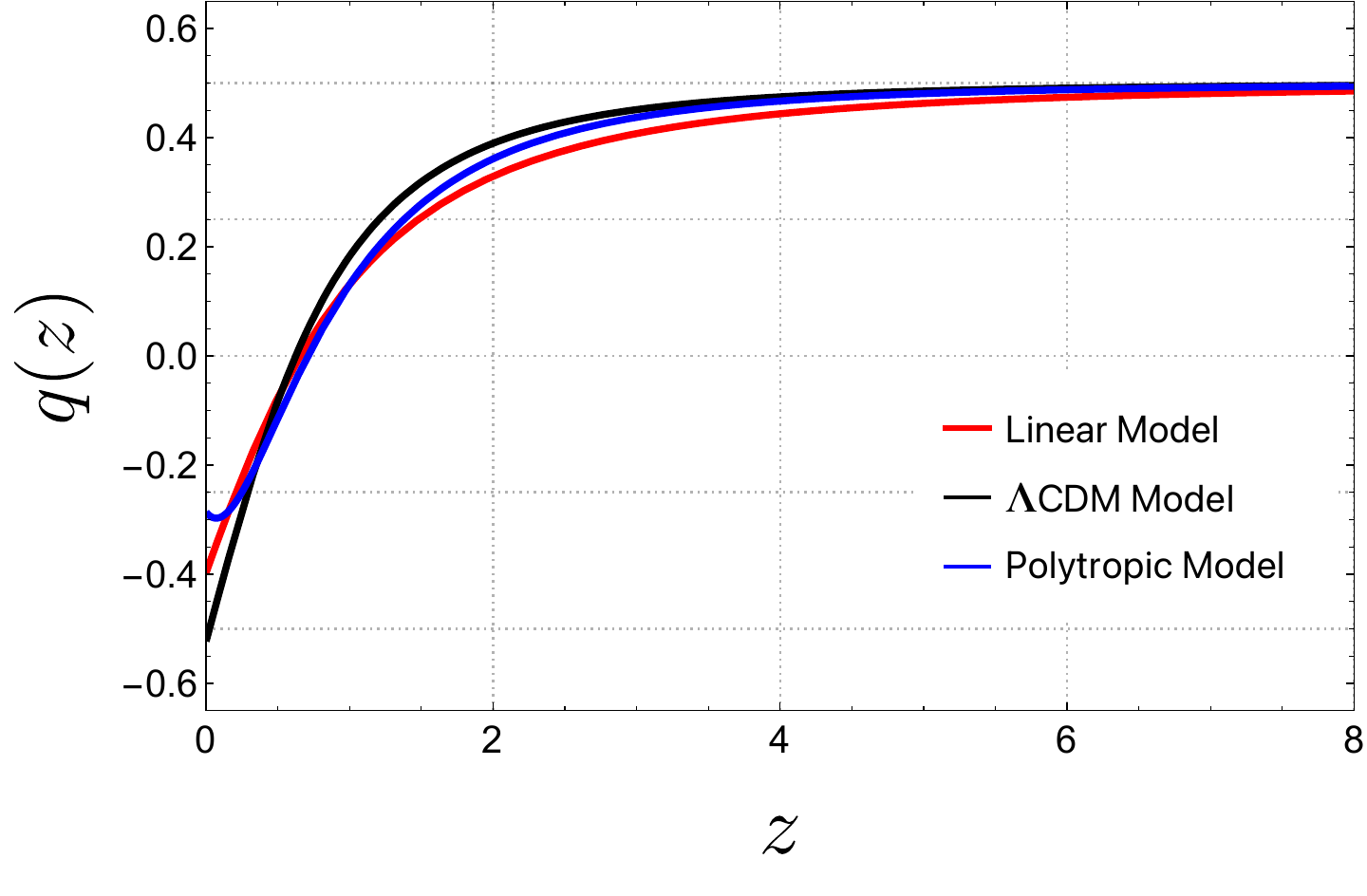}
\caption{Evolution of the deceleration parameter \( q(z) \) as a function of redshift \( z \), using the joint data best-fit values.}\label{fig_6}
\end{figure}
\subsubsection{Jerk parameter \( j(z) \)}
The jerk parameter \( j(z) \) \cite{jerksnap} measures the rate of change of the acceleration over redshift. It essentially tells us whether the acceleration is increasing, decreasing, or staying the same. Mathematically, it can be expressed as
\[
j(z) = \frac{1}{H^3(z)}\frac{d^2H(z)}{dz^2}.
\]
In the \(\Lambda\)CDM model, a key feature is that the jerk parameter is a constant value, namely $j(z)=1$ throughout the Universe's evolution. This constancy is a distinctive characteristic of \(\Lambda\)CDM, helping to differentiate it from other cosmological models where the jerk parameter might vary due to different underlying physics (as it also happens in our models).
\begin{figure}[H]
\centering
\includegraphics[width=8.5 cm]{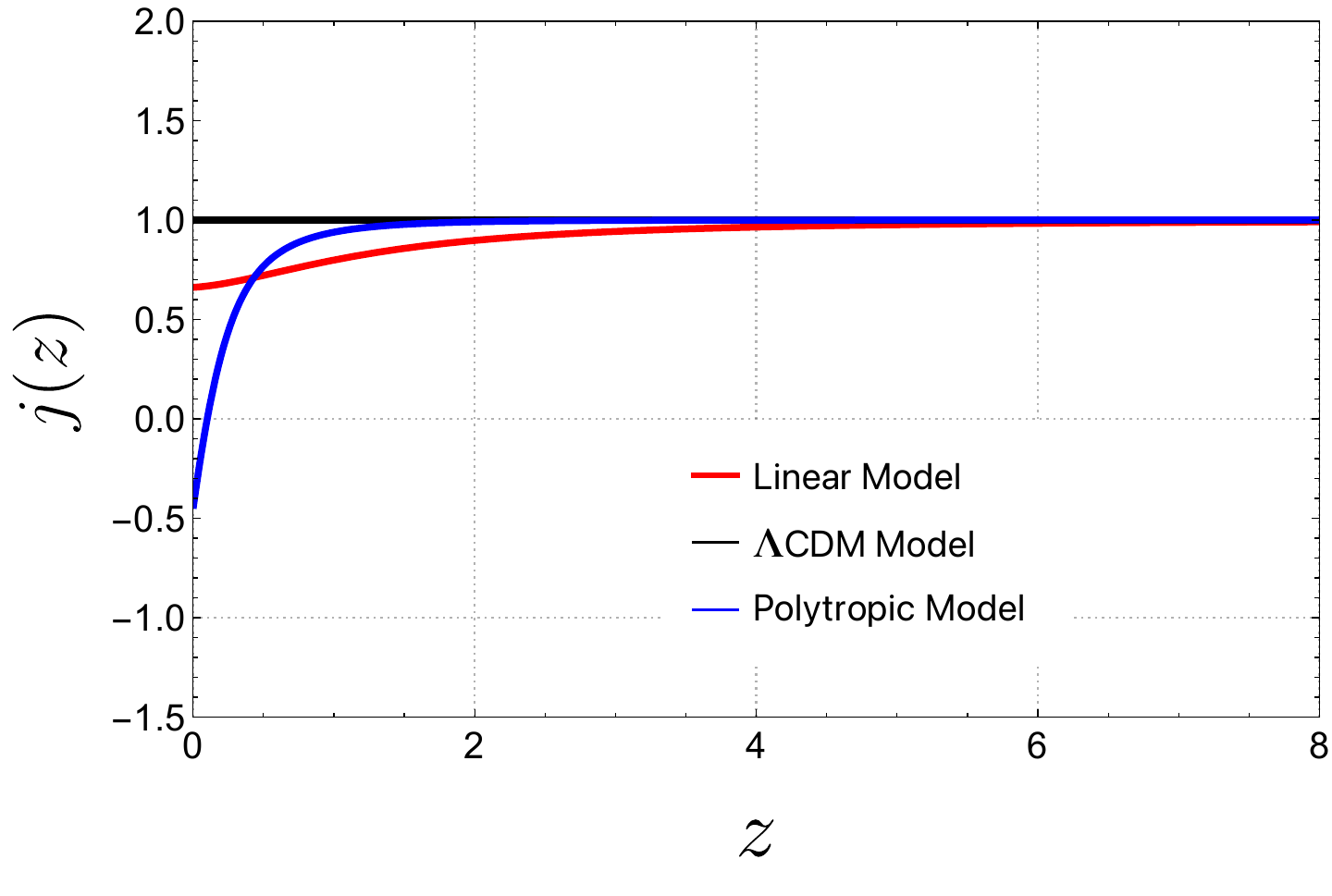}
\caption{Evolution of the jerk parameter \( j(z) \) as a function of redshift \( z \), using the joint data best-fit values}\label{fig_7}
\end{figure}
\subsubsection{Snap parameter \( s(z) \)}
The snap parameter $s(z)$ \cite{jerksnap} could differentiate between modified gravity models, in which the first two derivatives of the Hubble parameter agree. Essentially, it is related to the third derivative of the Hubble function, or formally 
\[
s(z) = \frac{1}{H^4(z)} \frac{d^3H(z)}{dz^3} = \frac{j(z) - 1}{3 \left( q(z) - \frac{1}{2} \right)}.
\]
From the second equality, we can see that it could be expressed through the jerk and deceleration parameters in a quite simple form. Though the snap parameter is less commonly referenced than the deceleration or jerk parameters, it plays a crucial role when testing more complex or modified cosmological models that may go beyond the standard \(\Lambda\)CDM framework. In the \(\Lambda\)CDM model, one of the defining features is that the snap parameter takes the constant value $0$, indicating that the rate of acceleration remains stable over redshift. This constancy of the snap parameter in \(\Lambda\)CDM helps distinguish it from other models that predict different dynamics.
\begin{figure}[H]
\centering
\includegraphics[width=8.5 cm]{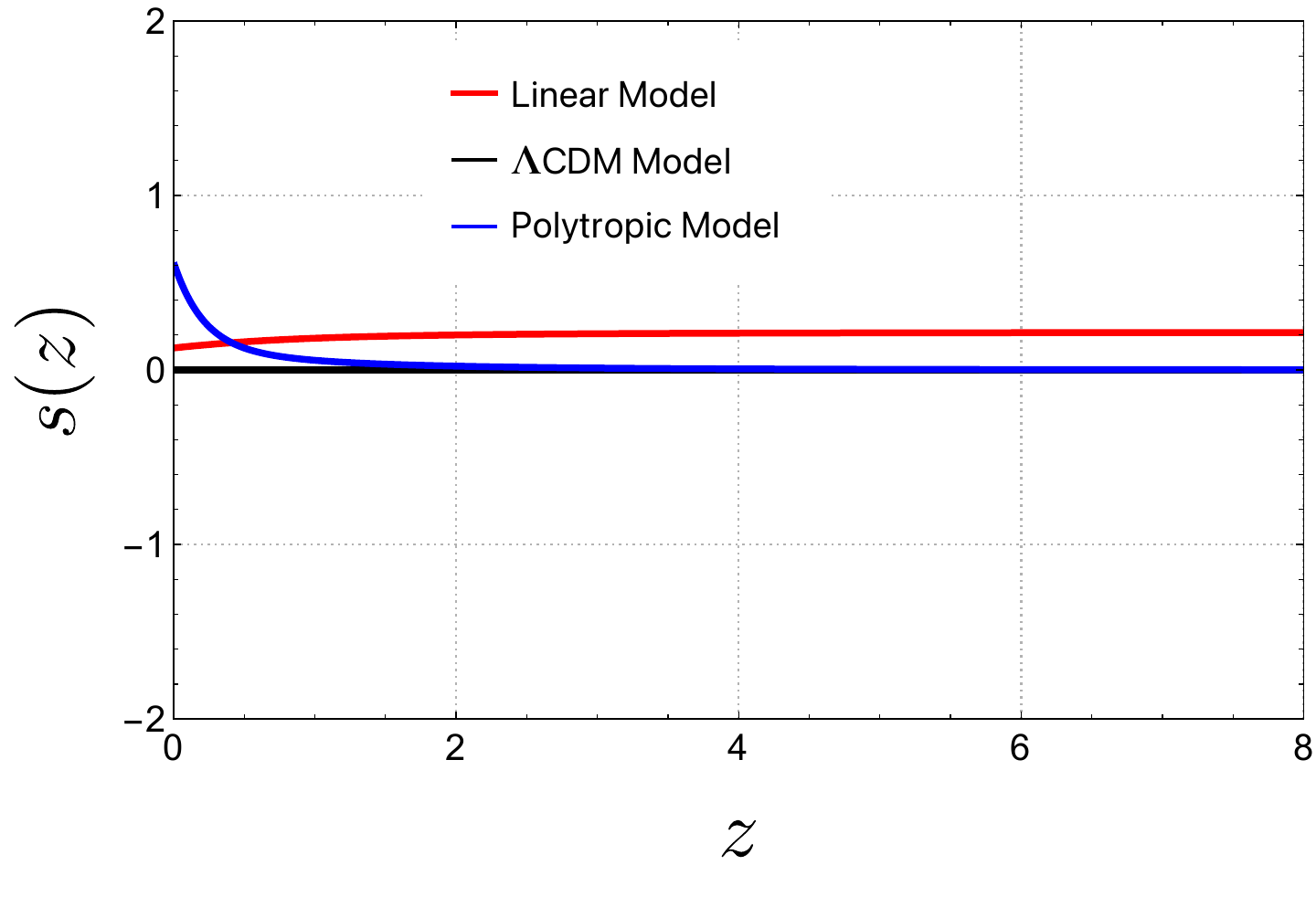}
\caption{Evolution of the snap parameter \( s(z) \) as a function of redshift \( z \), using the joint data best-fit values.}\label{fig_8}
\end{figure}
\subsection{$Om(z)$ diagnostic}\label{sec8}
To compare Semi-Symmetric Metric Gravity models with the standard \(\Lambda\)CDM model, we also use a key diagnostic tool known as the \( Om(z) \) diagnostic \cite{Om1,Om2,Om3,Om4}. This diagnostic is particularly useful for distinguishing between different cosmological models, which have quintessence-like, phantom-like evolutions. The \( Om(z) \) diagnostic is defined as
\[
Om(z) = \frac{H^2(z)/H_0^2 - 1}{(1 + z)^3 - 1} = \frac{h^2(z) - 1}{(1 + z)^3 - 1}.
\]
For the \(\Lambda\)CDM model, \( Om(z) \) is constant and equal to the present-day matter density, \( \Omega_{m0} = 0.322 \). This constancy reflects the model’s predictable and steady evolution. However, in alternative gravity theories that differ from \(\Lambda\)CDM, \( Om(z) \) may evolve over redshift. A changing \( Om(z) \) can reveal different types of cosmic behavior, as we mentioned before: if \( Om(z) \) increases with redshift (positive slope), it suggests a phantom-like evolution, where dark energy becomes increasingly dominant. On the other hand, if \( Om(z) \) decreases over time (negative slope), it indicates quintessence-like dynamics, where dark energy is less dominant, and the Universe’s expansion may slow down. If there are periods, in which the slope is positive, then it becomes negative, it indicates a transition from a phantom-like to a quintessence-like evolution.
\begin{figure}[H]
\centering
\includegraphics[width=8.5 cm]{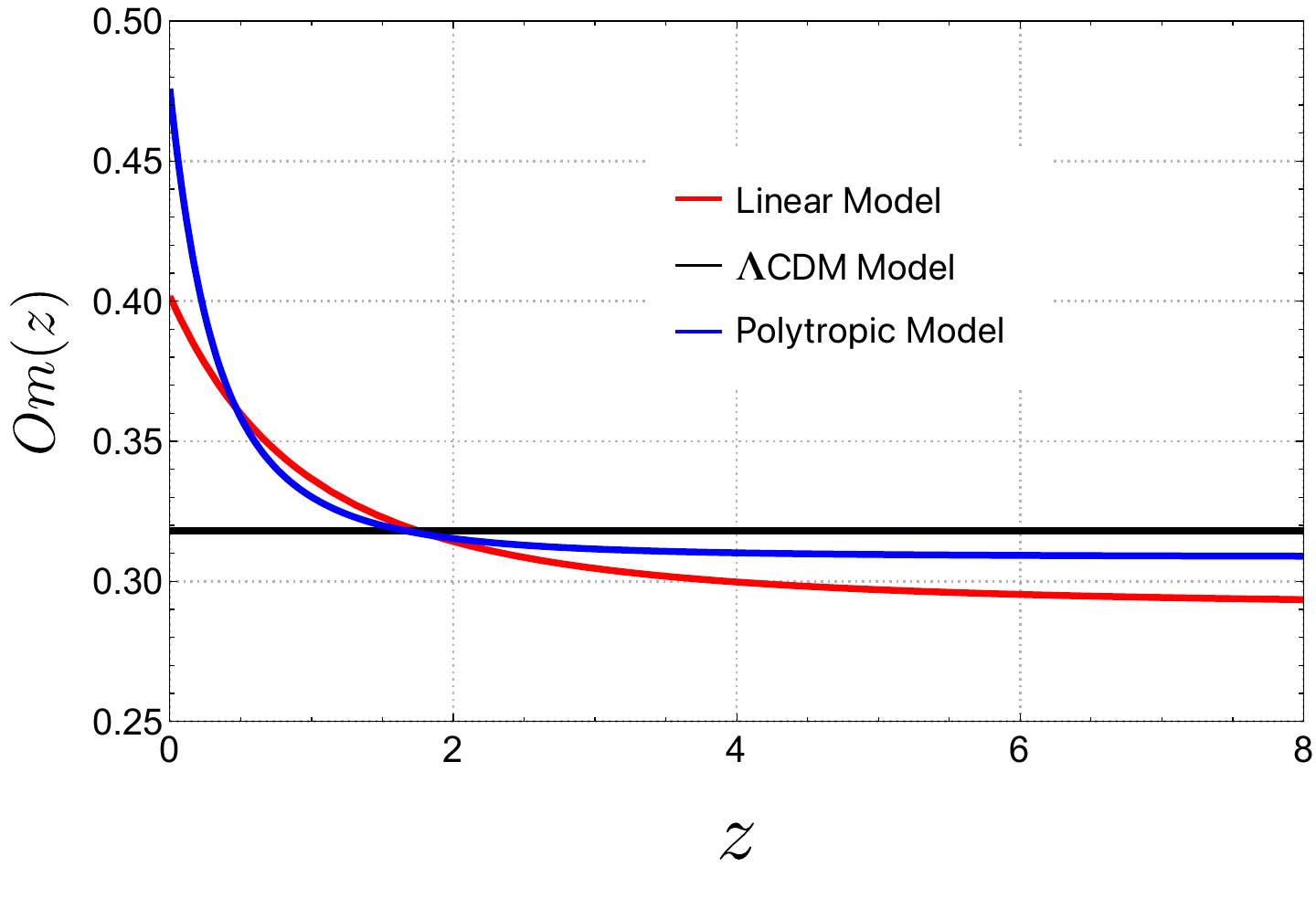}
\caption{Evolution of the \( Om(z) \) diagnostic as a function of redshift \( z \), using the joint data best-fit values.}\label{fig_9}
\end{figure}
\subsection{Matter density \( r(z) \) and the torsion vector \( \Omega(z) \)}
The dimensionless matter density represents an important cosmological parameter that allows a powerful check of the consistency of the cosmological models. In the present approach matter is considered together in both its forms (baryonic and dark), and its present day value determines the initial conditions for the cosmological evolution. In this sense the models do not predict the present day matter density, but give a description of its evolution in the earlier stages of evolution of the Universe. The torsion vector is a basic geometrical (and physical) component of the present models, and its cosmological evolution provides important hints on the nature of the torsion, and of the overall geometry of the Universe, as well as on the way it has evolved.   
\begin{figure}[H]
\centering
\includegraphics[width=8.5 cm]{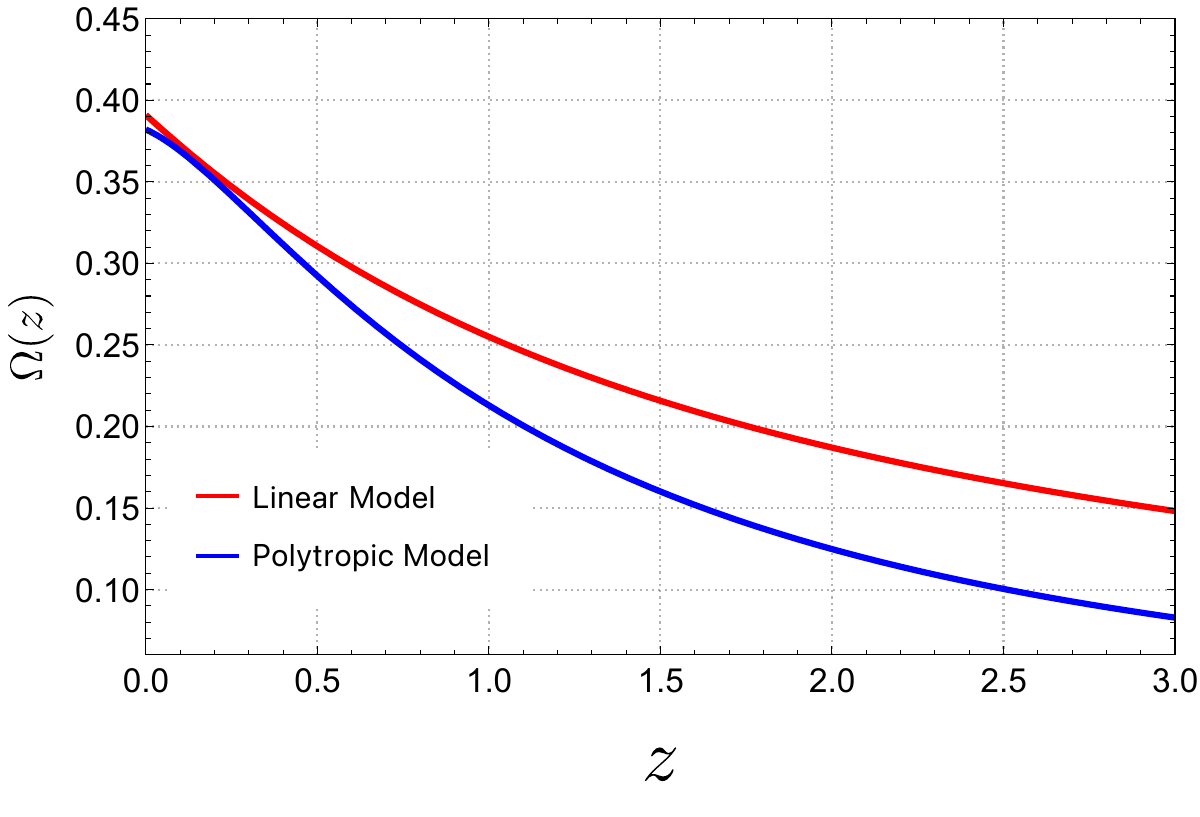}
\caption{Evolution of the dimensionless torsion vector \( \Omega(z) \) as a function of redshift \( z \), using the joint data best-fit values.}\label{fig_10}
\end{figure}
\begin{figure}[H]
\centering
\includegraphics[width=8.5 cm]{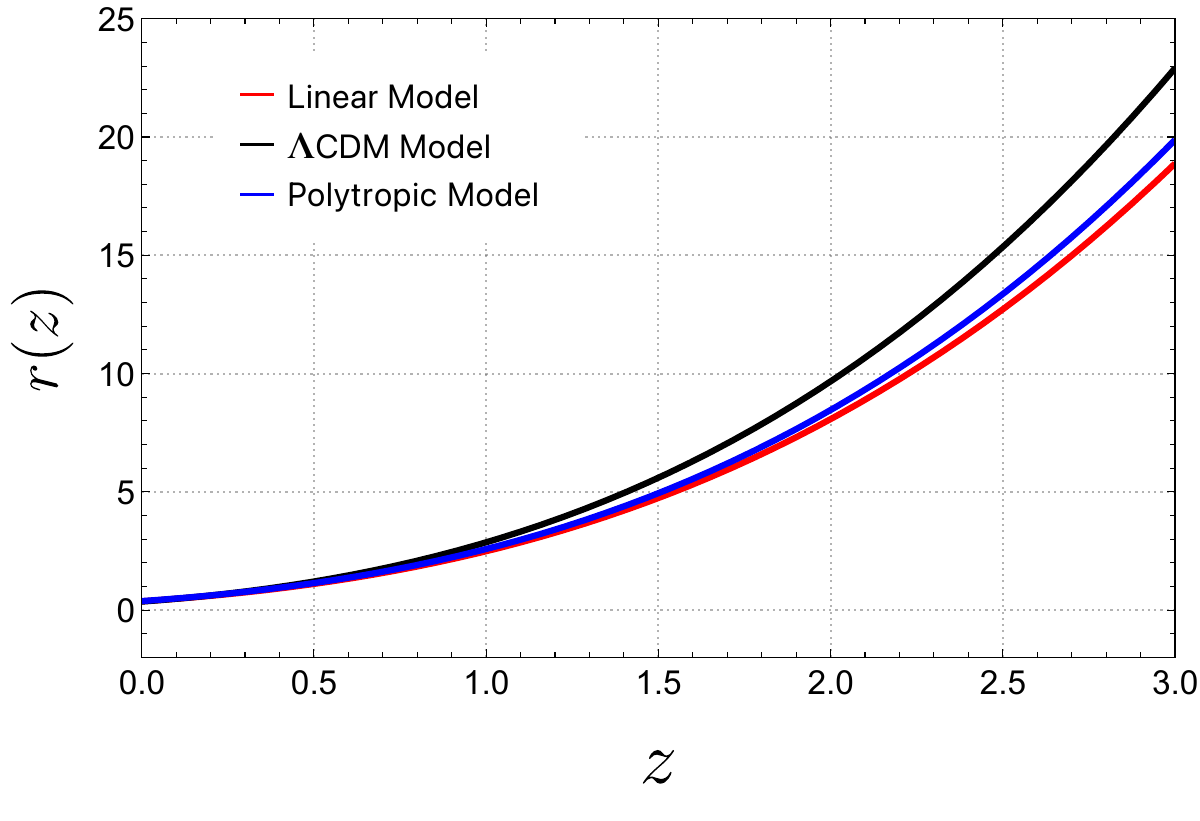}
\caption{Evolution of the dimensionless matter density \( r(z) \) as a function of redshift \( z \), using the joint data best-fit values.}\label{fig_11}
\end{figure}
\subsection{Particle creation rate $\Psi$ and creation pressure $P_c$.}
When the energy-momentum tensor is not conserved, as mentioned earlier, the particle number can vary. In our concrete case, the creation/annihilation processes are mostly influenced by the presence of the torsion vector. The quantities $P_c$ and $\Psi$ provide consistency conditions for the proposed models, as if the particle creation pressure would be positive, the models would not be consistent with the second law of thermodynamics. Thus, for a model to be valid, it must hold that $P_c <0 $ and $\Psi>0$ throughout the entire cosmic evolution.
\begin{figure}[H]
\centering
\includegraphics[width=8.5 cm]{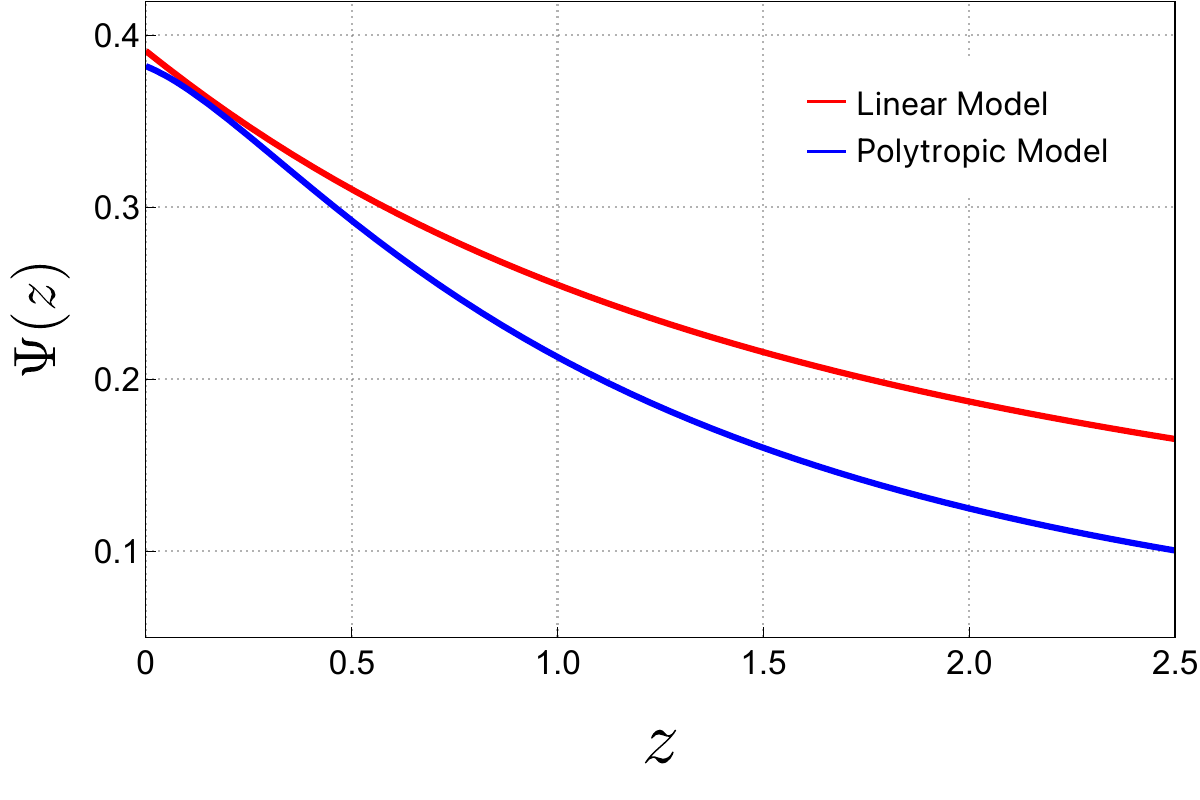}
\caption{Evolution of the dimensionless particle creation rate \( \Psi(z) \) as a function of redshift \( z \), using the joint data best-fit values.}\label{fig_creation}
\end{figure}
\begin{figure}[H]
\centering
\includegraphics[width=8.5 cm]{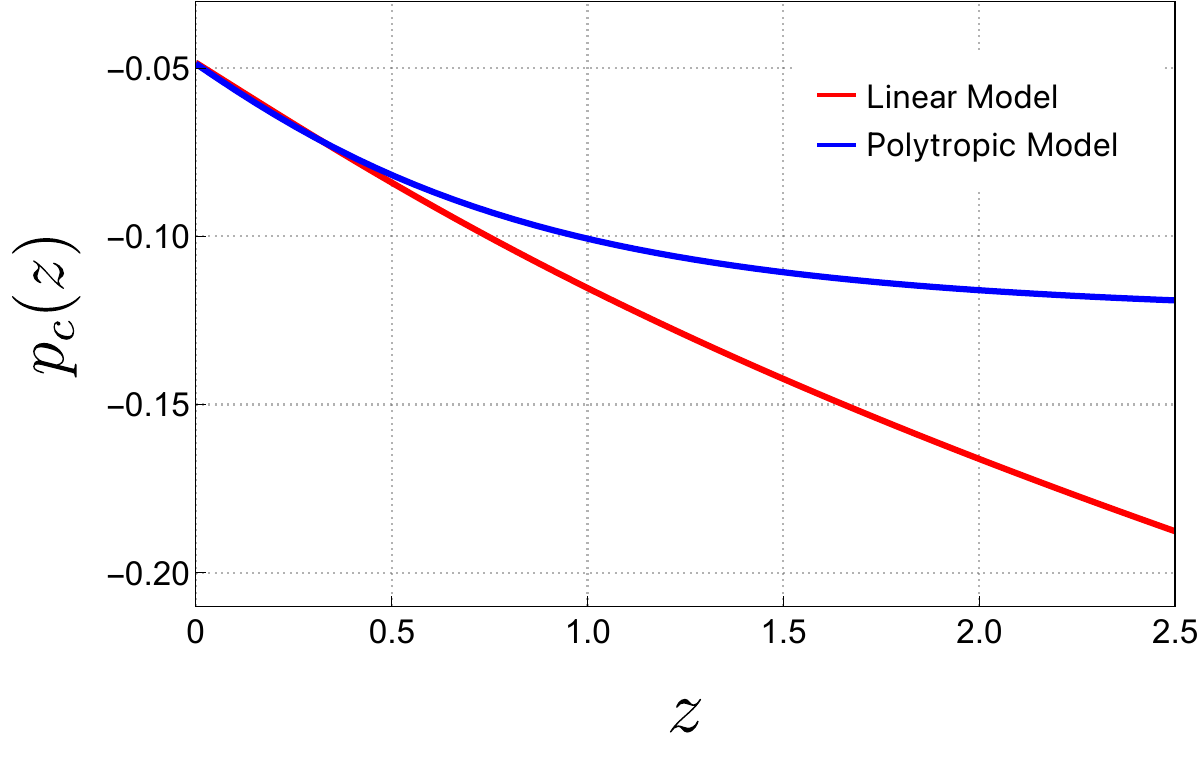}
\caption{Evolution of the dimensionless particle creation pressure \( P_c(z) \) as a function of redshift \( z \), using the joint data best-fit values.}\label{fig_pressure}
\end{figure}
\subsection{Statistical analysis}\label{sec9}
To differentiate between various models in the Semi-Symmetric Metric Gravity framework relative to the reference $\Lambda$CDM model, we begin by examining the minimum chi-squared values, \(\chi^2_{\text{min}}\), which quantify the fit quality for each model. The reduced chi-squared, $\chi_{\text{red}}^2$, is a statistical measure used to assess the goodness of fit for a model compared to the observed data. It is defined as
\begin{equation}
\chi_{\text{red}}^2 = \frac{\chi_{\text{tot}}^2}{\text{DOF}}
\end{equation}
where $\chi_{\text{tot}}^2$ is the total chi-squared value, and DOF (degrees of freedom) is the number of data points minus the number of fitted parameters. A value of $\chi_{\text{red}}^2 \approx 1$ suggests a good fit between the model and the data. If $\chi_{\text{red}}^2 < 1$, it might suggest overfitting. If $\chi_{\text{red}}^2 \gg 1$, it indicates a poor fit, meaning the model does not describe the data well. In addition to \(\chi^2_{\text{red}}\), we use statistical metrics derived from log-likelihood values to further evaluate the models. The maximum log-likelihood, \(\log L_{\text{max}} = \max(\log L_i)\), is obtained from the log-probability values and represents the maximum likelihood of the data being considered. In this analysis, the full dataset is used without incorporating the R22 prior. The Akaike Information Criterion (AIC) \cite{Liddle,AIC1,AIC2,AIC3,AIC4,AIC5} is then calculated as
\begin{equation}
\text{AIC} = -2 \log L_{\text{max}} + 2 P_{\text{tot}},
\end{equation}
where \(P_{\text{tot}}\) represents the number of free parameters in the model. The Bayesian Information Criterion (BIC) \cite{Liddle,BIC1} is computed using
\begin{equation}
\text{BIC} = -2 \log L_{\text{max}} + P_{\text{tot}} \ln(N_{\text{tot}}),
\end{equation}
with \(N_{\text{tot}}\) being the total number of data points. To assess the relative performance of each model, we calculate the differences in AIC and BIC with respect to the $\Lambda$CDM model:
\begin{equation}
\Delta \text{AIC} = \text{AIC}_{\text{model}} - \text{AIC}_{\Lambda \text{CDM}},
\end{equation}
\begin{equation}
\Delta \text{BIC} = \text{BIC}_{\text{model}} - \text{BIC}_{\Lambda \text{CDM}}.
\end{equation}
Additionally, the chi-squared value is derived from the log-probability values as: $\chi^2 = -2 \ln \mathcal{L}$. According to Jeffreys' scales, if \(0 < |\Delta \text{AIC}| \leq 2\), the models are considered comparable; if \(|\Delta \text{AIC}| \geq 4\), the model with the higher AIC is less favored. For BIC, \(0 < |\Delta \text{BIC}| \leq 2\) indicates weak disfavor, \(2 < |\Delta \text{BIC}| \leq 6\) indicates strong disfavor, and \(|\Delta \text{BIC}| > 6\) indicates very strong disfavor.
A negative value of \(\Delta \text{AIC}\) and \(\Delta \text{BIC}\) indicates that the model is preferred over the $\Lambda$CDM model.
\begin{table*}
\centering
\begin{tabular}{|l|c|c|c|c|c|c|c|c|}
    \hline
    \textbf{Models} & \textbf{${\chi_{\text{tot},min}^2}$} & \textbf{$P_{tot}$} & \textbf{$N_{tot}$} & \textbf{$\chi_{\text {red }}^2$} & \textbf{AIC} & \textbf{$\Delta$AIC} & \textbf{BIC} & \textbf{$\Delta$BIC} \\ 
    \hline\hline
    $\Lambda$CDM & 1801.11 & 4 & 1758 & 1.026 & 1809.11 & 0 & 1831.00 & 0 \\ 
    \hline
    Linear & 1790.07 & 5 & 1758 & 1.021  & 1800.07 & -9.04 & 1827.43 & -3.57 \\ 
    \hline
    Polytropic & 1787.79 & 5 & 1758 & 1.019 & 1797.79 & -11.32 & 1825.15 & -5.85 \\ 
    \hline
\end{tabular}
\caption{Summary of ${\chi_{\text{tot},min}^2} $, $\chi_{\text {red }}^2$, AIC, $\Delta$AIC, BIC, $\Delta$BIC for the $\Lambda$CDM Model, Linear, and Polytropic Models}
\label{tab_3}
\end{table*}
\subsection{Results}

\subparagraph{Description of best-fit values and triangle plot.}
Figs.~\ref{fig_1} and ~\ref{fig_2} show the triangle plot, also known as a corner plot, which is a visualization tool frequently used in Bayesian statistics. These plots correspond to the $\Lambda$CDM, Linear and Polytropic models, respectively. They provide a compact way to display the posterior distributions of multiple parameters and their correlations. The diagonal Elements (1D Histograms) are the marginalized posterior distributions for individual parameters. They give insights into the spread and peak of the distribution, indicating the most probable values for each parameter. The Off-Diagonal Elements (2D Contours)  plots represent the joint probability distributions between pairs of parameters. The contours visualize the correlation between two parameters, showing how the uncertainty in one parameter might influence the other. The Table~\ref{tab_2} presents a comparative analysis of the optimal values for cosmological parameters within the framework of Semi-Symmetric Metric Gravity across two models: the Linear Model, and the Polytropic Model. A key focus is on the Hubble constant \( H_0 \) and the sound horizon at drag epoch \( r_d \), as these parameters are critical in understanding the Universe's expansion rate and the scale of baryon acoustic oscillations (BAO).  In the \(\Lambda\)CDM framework, the estimated values of the Hubble constant \(H_0\) and the sound horizon \(r_d\) are aligned with those reported by Planck 2018 \cite{Planck}. When incorporating the R22 prior for \(H_0\), the fit gives \(H_0 = 71.48 \pm 0.87 \, \mathrm{km} \, \mathrm{s}^{-1} \, \mathrm{Mpc}^{-1}\), which is consistent with the value reported by the SH0ES collaboration \cite{R22}. This model also estimates the sound horizon as \(r_d = 139.2 \pm 1.8 \, \mathrm{Mpc}\), which is lower than the Planck value. The Linear Model predicts \(H_0 = 66.9 \pm 1.6 \, \mathrm{km} \, \mathrm{s}^{-1} \, \mathrm{Mpc}^{-1}\), closely matching the Planck estimate. It also yields \(r_d = 146.8 \pm 3.4 \, \mathrm{Mpc}\), consistent with values reported in studies such as \cite{Lemos}. When the R22 prior is included, the Linear Model's Hubble constant shifts to \(H_0 = 71.28 \pm 0.89 \, \mathrm{km} \, \mathrm{s}^{-1} \, \mathrm{Mpc}^{-1}\), again aligning with the SH0ES result. However, the corresponding sound horizon decreases to \(r_d = 138.8 \pm 1.9 \, \mathrm{Mpc}\), differing from the Planck prediction. Similarly, the Polytropic Model, without the R22 prior, finds \(H_0 = 66.8 \pm 1.6 \, \mathrm{km} \, \mathrm{s}^{-1} \, \mathrm{Mpc}^{-1}\), in agreement with the Planck value. It estimates the sound horizon at \(r_d = 146.8 \pm 3.4 \, \mathrm{Mpc}\), consistent with previous studies, including \cite{Lemos} and \cite{Nunes}. When the R22 prior is added, the numerical value of the Hubble constant increases to \(H_0 = 71.25 \pm 0.88 \, \mathrm{km}\, \mathrm{s}^{-1}\, \mathrm{Mpc}^{-1}\), aligning with the SH0ES measurement, while the sound horizon decreases to \(r_d = 138.6 \pm 1.9 \, \mathrm{Mpc}\), further diverging from the Planck value. These results highlight the persistent tension between early- and late-Universe measurements of \(H_0\) and \(r_d\). The values of \(H_0\) inferred from models using the R22 prior align with local measurements by SH0ES but deviate from the Planck-based inference, which relies on early-Universe observations. Conversely, the corresponding estimates of \(r_d\) in models incorporating R22 are systematically smaller than the Planck values, further reinforcing the discrepancy between early and late-Universe cosmological parameters. The Linear and Polytropic models use \(\Omega_0\) as the initial condition for their differential equation systems. For the Linear Model, \(\Omega_0\) is estimated to be \(0.3914 \pm 0.0095\), while for the Polytropic Model, \(\Omega_0\) is estimated to be \(0.3808 \pm 0.0083\). In the Linear Model, the free parameter \(\sigma_0\) is estimated to be \(0.950 \pm 0.047\), while in the Polytropic Model, the parameter \(K\) is estimated to be \(-1.91_{-0.15}^{+0.17}\). The absolute magnitude of Type Ia supernovae, \(\mathcal{M}\), remains consistent across all models, with values ranging from \(-19.424 \pm 0.055\) to \(-19.453 \pm 0.053\).\\\\
\subparagraph{Hubble Parameter \( H(z) \).}
The comparative analysis between the \( \Lambda \)CDM paradigm and each model against the CC data is presented in Fig~\ref{fig_3}. The \( \Lambda \)CDM model is shown as a black line, the Linear model as a red line, and the Polytropic model as a blue line. The CC measurements are represented by blue dots with magenta error bars. Notably, for redshifts \( z > 1.75 \), a clear deviation between the models emerges. Although these deviations are not substantial for low redshifts, they become apparent as redshift increases. However, for lower redshifts \( z < 1.75 \), all models exhibit close agreement with  each other and the observational data.\\\\ 
\subparagraph{Hubble difference \( \Delta H(z) \).}
Fig~\ref{fig_4} shows the Hubble difference \( \Delta H(z) \) between the Linear, Polytropic, and \( \Lambda \)CDM models compared to cosmic chronometer (CC) measurements. The plot illustrates the deviations of both model from the \( \Lambda \)CDM model at redshifts \( z > 1.5 \). These deviations are relatively minor, and as the redshift decreases to \( z < 1.5 \), the deviations become less pronounced, and all models align more closely with the observational data in this regime.\\\\
\subparagraph{Distance Modulus \( \mu(z) \).}
Fig~\ref{fig_5} illustrates the evolution of \( \mu(z) \) as a function of redshift. Both models exhibit close agreement with the Type Ia supernova (SNe Ia) measurements and the $\Lambda$CDM Model. Although the differences are subtle and may not be immediately apparent due to the large number of data points, we have included an additional figure to highlight the variations more clearly. This additional Figure demonstrates the deviations between the $\Lambda$CDM, Linear, and Polytropic Models, allowing for a more detailed comparison of their respective evolutions.\\\\
\subparagraph{Cosmographic results.}
Fig.~\ref{fig_6} illustrates the evolution of the deceleration parameter \( q(z) \). At high redshifts, the \(\Lambda\)CDM, Linear, and Polytropic models show similar behavior. As redshift decreases toward the present day (\( z = 0 \)), the \(\Lambda\)CDM model predicts \( q_{0} = -0.506 \), the Linear model \( q_{0} = -0.401 \), and the Polytropic model \( q_{0} = -0.288 \). The redshift \( z_{tr} \), where \( q(z) = 0 \), represents the point at which the expansion of the Universe transitions from decelerating to accelerating. The \(\Lambda\)CDM model predicts this transition at \( z_{tr} = 0.601 \), implying that this shift occurred relatively recently. In contrast, the Linear model predicts a later transition at \( z_{tr} = 0.642 \), while the Polytropic model forecasts an even later transition at \( z_{tr} = 0.711 \). Fig~\ref{fig_7} shows the redshift variation of the jerk parameter \( j(z) \) with respect to redshift \( z \). At high redshifts, the $\Lambda$CDM, Linear and Polytropic models align closely, all exhibiting a value of \( j(z) = 1 \), which means that the jerk parameter of our models asymptotically tends to the constant value $1$, but for lower redshifts, we have different dynamics.  The $\Lambda$CDM model  predicts \( j(z) = 1 \) throughout the entire evolution, including at \( z = 0 \). In comparison, the Linear model predicts \( j_0 = 0.662 \), while the Polytropic model predicts \( j_0 = -0.432 \). The value of \( j_0 \) at \( z = 0 \) is particularly important because it characterizes the present-day cosmic expansion. The fact that the $\Lambda$CDM model maintains \( j_0 = 1 \) reinforces its consistency with a Universe dominated by a cosmological constant, while the deviation of \( j_0 \) in the other models suggests alternative dynamics in the current expansion rate, driven by torsion. Fig~\ref{fig_8} depicts the dependence of the snap parameter \( s(z) \) on to redshift \( z \). At high redshifts, the $\Lambda$CDM and Polytropic models are closely aligned, both exhibiting \( s(z) = 0 \). In contrast, the Linear model shows a slight deviation, with \( s(z) = 0.228 \) at high redshifts. Although this deviation is minor, it is still noticeable.  At lower redshifts, particularly at \( z = 0 \), the present-day value of the snap parameter, \( s_0 \), is of particular interest as it characterizes the current cosmic expansion. The $\Lambda$CDM model predicts \( s_0 = 0 \), reflecting its consistent behavior throughout the evolution of the Universe. In contrast, the Linear model predicts \( s_0 = 0.143 \), while the Polytropic model predicts \( s_0 = 0.602 \).\\\\
\subparagraph{$Om(z)$ Diagnostic.}
Fig ~\ref{fig_9} presents the evolution of the $Om(z)$ diagnostic as a function of the redshift. In both the linear and polytropic models, $Om(z)$ exhibits a monotonic decrease with increasing redshift. This behavior is characteristic of a quintessence-like evolution.\\\\
\subparagraph{Variation of matter density and of the torsion vector.}
Figs~\ref{fig_10} and \ref{fig_11}, which depict the evolution of $\Omega(z)$ and $r(z)$ as a function of redshift. In both models, the matter density predictions align with those of the $\Lambda\text{CDM}$ model up to $z \approx 1$. However, at higher redshifts, both models (Linear and Polytropic) predict a lower matter density compared to the standard $\Lambda\text{CDM}$ model. Notably, the Polytropic Model predicts a higher matter density than the Linear Model at these higher redshifts. On the other hand, in both models, the torsion vector decreases monotonically as the redshift increases.\\\\
\subparagraph{Particle creation rate $\Psi$ and Creation Pressure $P_c$.}
On Figure \ref{fig_creation} the dimensionless particle creation rate $\Psi(z)$ is illustrated for the two proposed cosmological models. In both cases, the creation rate is positive, and monotonically decreasing as a function of redshift throughout the evolution. The numerical values of $\Psi(z)$ for the two models are close to each other  for small redshifts $0<z<0.5$, and a more substantial difference can only be observed at higher redshifts. As the creation rate does not cross zero for either of the models, particle annihilation processes do not take place. This is in accordance with the second law of thermodynamics, and with the fact that the creation pressure, as depicted on Fig.~\ref{fig_pressure} takes negative values throughout the evolution. The pressure of the created particles is also monotonically decreasing as a function of $z$, hence reaching its maximum value at $z=0$. The linear polytropic model shows a tendency to converge to a constant negative creation pressure for large redshifts ($z>2)$, while in the linear model the pressure keeps heavily decreasing. \\\\
\subparagraph{Statistical results.}
Table ~\ref{tab_3} offers a detailed comparison of the $\Lambda$CDM model, the Linear Model, and the Polytropic Model based on several key metrics: minimum chi-squared (\(\chi^2_{\text{min}}\)), reduced chi-squared (\(\chi_{\text{red}}^2\)), Akaike Information Criterion (AIC), Bayesian Information Criterion (BIC), and their respective differences (\(\Delta\text{AIC}\) and \(\Delta\text{BIC}\)). The (\(\chi^2_{\text{min}}\)) values for the models are 1801.11 for the $\Lambda$CDM model, 1790.07 for the Linear model, and 1787.79 for the Polytropic model. The reduced chi-squared (\(\chi_{\text{red}}^2\)) values, which account for the degrees of freedom, are quite similar across the models: 1.026 for $\Lambda$CDM, 1.021 for Linear, and 1.019 for Polytropic. Since all the \(\chi_{\text{red}}^2\) values are close to 1, this suggests that each model fits the data reasonably well. The Polytropic model, with the lowest \(\chi_{\text{red}}^2\) value, indicates a slightly better fit compared to the others, though the difference is minimal. All models show reduced chi-squared values close to 1, indicating that they fit the data reasonably well, with the Polytropic model having the smallest reduced chi-squared value, suggesting a marginally better fit among the three. The AIC values are 1809.11 for the $\Lambda$CDM model, 1800.07 for the Linear model, and 1797.79 for the Polytropic model. The differences in AIC (\(\Delta \text{AIC}\)) relative to the $\Lambda$CDM model are -9.04 for the Linear model and -11.32 for the Polytropic model.  Negative values of \(\Delta \text{AIC}\) indicate that both the Linear and Polytropic models are preferred over the $\Lambda$CDM model according to AIC, with the Polytropic model being the most favored. The BIC values are 1831.00 for the $\Lambda$CDM model, 1827.43 for the Linear model, and 1825.15 for the Polytropic model. The differences in BIC (\(\Delta \text{BIC}\)) relative to the $\Lambda$CDM model are -3.57 for the Linear model and -5.85 for the Polytropic model. According to BIC, the Polytropic model is again preferred over the Linear and $\Lambda$CDM models, with the largest \(\Delta \text{BIC}\) suggesting the strongest support for the Polytropic model. While all three models exhibit similar reduced chi-squared values, indicating similar goodness of fit, the AIC and BIC metrics reveal a preference for the Linear and Polytropic models over the $\Lambda$CDM model. Among these, the Polytropic model shows the most favorable results in both AIC and BIC, suggesting that it provides the best balance between fit quality and model complexity. The Linear model also performs well but is slightly less favored compared to the Polytropic model according to both criteria.\\\\
\section{Discussions and final remarks}\label{sect3}
In the present paper we have briefly reviewed an interesting geometric extension of general relativity, based on a concept of torsion that was introduced one hundred years in mathematics by Friedmann and Schouten \cite{FS}. In this approach torsion is essentially a vectorial quantity, and it is described by a single four-vector $\pi_\mu$, a description that significantly reduces the number of the torsion components to four. Moreover, the torsion tensor takes the very simple and elegant form of Eq.~(\ref{semisymmetric}), which also leads to a significant decrease in the calculation complexity of the geometry, and of the physical models. By analogy with the Einstein equations, we have postulated a set of field equations, in which the Einstein tensor, constructed in the Semi-Symmetric geometry, is proportional to the matter energy-momentum tensor. In this way we obtain a set of gravitational field equations in which the torsion tensor, as well as its derivatives, explicitly appear as a geometric contribution, expressed in terms of a single vector field. The extra terms thus generated geometrically can be interpreted physically as corresponding to dark energy, dark matter, or both. In the present investigation we have considered that the extra geometric part of the field equations can be interpreted only as dark energy, and we have assumed the existence of a single matter component, consisting of both dark and baryonic matter. 

The Newtonian limit of the relativistic gravitational models offers not only the possibility of comparing the theory with the Newtonian gravity, and also provides some important possibilities of testing its predictions. We have obtained the corrections to the Newtonian potential induced by the presence of the torsion. as well as the corrections to Newton's law of force, in the presence of a torsion vector having only a radial nonzero component $\Pi_r$. The $1/r^2$ terms in Newton's law are corrected by a term proportional to $\exp \left(\Pi_r r\right)$, and by a term of the form $-4/\Pi_r r^2$, respectively. A term proportional to $\Pi_r$, as well as one proportional to $-1/r$ also do appear in the gravitational force equation. These terms could induce some physical effects at the Solar System level that could allow the testing of the presence of torsion in the Universe. However, the role of these terms may be dominant at the galactic level, and they could provide an explanation for the behavior of the galactic rotation curves without any need for the presence of a dark matter component.

We have also investigated in detail the cosmological implications of the Semi-Symmetric Metric Gravity models, by considering two simple models obtained by imposing some specific equations of state for the effective geometric dark pressure. The presence of the torsion naturally generates supplementary terms in the cosmological Friedmann equations, which can be interpreted as an effective dark energy and pressure. By imposing some relations between these two quantities, representing a specific equation of state, we can build consistent cosmological models, which can describe well the observational data at least in the redshift range $z\in (0,3)$.  All models describe an initially decelerating phase, followed by a transition to an accelerated one. 

The possible presence of torsion in the Universe has been extensively investigated in various theoretical frameworks. In particular, the Einstein-Cartan type theories \cite{Hehl, EC1,EC2} have attracted a lot of attention. In the Einstein-Cartan theory torsion is directly related to a physical property of matter, with the torsion proportional to the spin tensor. Hence, in these models, rotation is the source of the torsion. On the other hand, in the present approach of the Semi-Symmetric Metric Gravity, torsion is interpreted as an intrinsic property of the spacetime, coexisting together with the metric as an independent variable. 

Of particular interest in the Friedmann-Schouten geometry is the form of the torsion tensor, which is purely vectorial. This leads to a significant simplification of the mathematical formalism, with the field equations taking a (relatively) easily tractable form. This form of the torsion does appear naturally (in its specific mathematical framework), and it is not related to any particular condition imposed on the tensor $T^\mu _{\;\;\nu \rho}$. For example, in \cite{Ios}, it was shown that the torsion tensor can be decomposed as $T_{abc}=\phi \left(h_{ab}u_c-h_{ac}u_b\right)$, where $u_c$ is the four-velocity, and $h_{ab}=g_{ab}+u_au_b$.  The generalized Friedmann equations of the Semi-Symmetric Metric Gravity theory contain the extra contributions coming from the torsion vector, and they naturally lead to the possibility of building dark energy models that could give a good description of the observational data. 

One of the interesting features of the present theory is that it does not introduce novel coupling parameters, and arbitrary coefficients. The dynamical evolution of the Universe is determined only by the present day values of the torsion vector, fully determined by the matter density, which is known from observations. Hence, the entire cosmological dynamics can be described with the help of a single parameter, the value at $z=0$ of the dimensionless function $r$.         

In our analysis we have restricted our investigations to the case of the flat FLRW geometry. Curvature effects may play an important theoretical role in the cosmological evolution of the considered models. The effect of the spatial curvature $k$ in the evolution of the Universe is described by the curvature density parameter $\Omega_k$, defined according to $\Omega _k=-kc^2/a^2H^2$ \cite{NF}. $\Omega_k$ can take positive, zero or negative values, depending on the values of $k$, $k=-1,0,+1$, corresponding to open, flat, or closed geometries. 

The cosmic curvature density parameter $\Omega_k$  has been constrained in \cite{NF} independently  of any background cosmological model, by adopting the nonparametric Gaussian processes, with the only assumption that the Universe is homogeneous and isotropic, and described by the FLRW metric. The results of the investigations in \cite{NF} show that a spatially flat Universe is consistent in 2$\sigma$ within the redshift range $0<z<2$ for the background data.  By also taking into account the Red Shift Distorsion (RSD) data, the results are consistent with a spatially flat Univere mostly within 2$\sigma$, and always within 3$\sigma$ in the redshift range $0<z<2$. 
These results indicate that the assumption of the geometric flatness, implemented via the $\Lambda$CDM model, is well supported by the observational data.

 Since the models we have considered are very close to the $\Lambda$CDM paradigm, and even  reproducing it almost exactly for a specific range of model parameter values, the effects of the curvature density parameter $\Omega _k$ can be ignored in our analysis that extends also only to the redshift range $0<z<2$.

 On the other hand, at higher redshifts, curvature effects may have a more important impact on the overall cosmological evolution, even that the inflationary evolution is expected to drastically reduce the presence of curvature.   

The analysis of the cosmological models of the Semi-Symmetric Metric Gravity theory  has been done under the simplifying assumption of a Universe filled with a perfect cosmological fluid. But in many astrophysical and cosmological configurations the idealized perfect fluid model of matter may not be suitable, especially in the case of matter at very high densities and pressures. 

Such possible situations are the classical description of the (quantum) particle production phase, quark and gluon plasma viscosity, mixtures of cosmic elementary particles, evolution of cosmic strings due interaction with matter, relativistic transport of photons, interaction between matter and radiation etc. (see \cite{Visc} and references therein).

There are two basic theories of bulk viscosity, the Eckart and the Israel-Stewart theories \cite{Visc1}. The bulk viscosity effects can be generally described by means of an effective pressure $p_{bulk}$, which can be formally added to the thermodynamic pressure to generate an effective pressure  $p_{eff}=P+p_{bulk}$ \cite{Visc1}, leading to an effective energy-momentum tensor with components $T_0^0=\rho$, $T_i^i=-p_{eff}$, $i=1,2,3$.

 In the simplest approach to bulk viscous phenomena, the Eckart theory, and in the cosmological context, the bulk viscous pressure is assumed to have the form $p_{bulk}=-3\xi H$, where $\xi$ is the coefficient of the bulk viscosity. Hence, in the case of matter creation in the presence of bulk viscosity, in the comoving frame the energy-momentum tensor has the components  $T_0^0=\rho$, $T_i^i=-\left(P+P_c-3\xi H\right)$, $i=1,2,3$. As one can see from the form of the energy-momentum tensor, the bulk viscous pressure acts as a creation pressure, with the opposite interpretation also possible, the creation pressure being related to an effective viscosity of the medium in which particle creation takes place.

 As for the energy balance equation for a viscous fluid, and in the presence of matter creation, it takes the form
\begin{equation}
\dot{\rho}+3H\left(\rho+P+P_c\right)=3\xi H^2,
\end{equation} 
showing again that the bulk viscous term can also be interpreted as describing an effective particle creation process, even in the absence of the creation pressure introduced via the thermodynamics of irreversible processes. 

In the present approach we have considered that the newly created matter behaves as an ideal fluid, and we have neglected the viscous effects. This assumption is physically motivated by the fact that we expect the particle creation effects to generate matter at a very low density and pressure, and thus the bulk viscous effects can be considered as negligibly small.

Another possible application of the Semi-Symmetric Metric Gravity would be to consider inflation in the presence of torsion, an approach that could leas to a novel perspective
on the cosmological, gravitational and astrophysical processes that significantly influenced and modelled the early dynamics of the Universe after the Big Bang. 

Consequently, the predictions of the present model could introduce major differences, as compared to those of standard general relativity, or its extensions and generalizations that ignore the role of torsion. These differences could also have a significant impact on several current areas of present day physics, which are intensively investigate, including astrophysics, cosmology, the study of the gravitational collapse, and the properties of the gravitational waves. 

To conclude, in the present investigation, we have introduced a novel torsion-based gravitational theory, and we have explicitly shown its
theoretical and observational consistency. This novel approach also encourages, and strongly motivates, the theoretical investigation of further extensions of the families of gravitational theories that go beyond the  framework of the standard Riemannian geometry. 

\section*{Funding}
This research received no external funding

\section*{Institutional Review Board Statement}
Not applicable.

\section*{Informed Consent Statement}
Not applicable.

\section*{Data Availability Statement}
The data underlying this article is already given with references during the analysis of this work.

\section*{Acknowledgments}
We would like to thank the three anonymous reviewers for comments and suggestions that helped us to significantly improve our manuscript, and to Sunny Vagnozzi for fruitful discussions. L.Cs. would like to express his gratitude to the Collegium Talentum Programme of Hungary.

\section*{Conflicts of Interest }
The author declares no conflicts of interest. The funders had no role in the design of the study; in the collection, analyses, or interpretation of data; in the writing of the manuscript; or in the decision to publish the results

\end{document}